\documentclass{emulateapj}
\submitted{{\it Submitted for publication in ApJ}}
\usepackage{multirow,color,wrapfig,ulem}
\usepackage {graphicx}
\usepackage{graphics}
\usepackage[dvips]{epsfig}

\newcommand{\ly}{{\ifmmode{{\rm Ly}\alpha~}\else{Ly$\alpha$~}\fi}}
\newcommand{\hMpc}{{\ifmmode{h^{-1}{\rm Mpc}}\else{$h^{-1}$Mpc }\fi}}   
\newcommand{\hGpc}{{\ifmmode{h^{-1}{\rm Gpc}}\else{$h^{-1}$Gpc }\fi}}   
\newcommand{\hmpc}{{\ifmmode{h^{-1}{\rm Mpc}}\else{$h^{-1}$Mpc }\fi}}  
\newcommand{\hkpc}{{\ifmmode{h^{-1}{\rm kpc}}\else{$h^{-1}$kpc }\fi}}  
\newcommand{\hMsun}{{\ifmmode{h^{-1}{\rm
        {M_{\odot}}}}\else{$h^{-1}{\rm{M_{\odot}}}$}\fi}}   
\newcommand{\hmsun}{{\ifmmode{h^{-1}{\rm
        {M_{\odot}}}}\else{$h^{-1}{\rm{M_{\odot}}}$}\fi}}   
\newcommand{\Msun}{{\ifmmode{{\rm {M_{\odot}}}}\else{${\rm{M_{\odot}}}$}\fi}}  
\newcommand{\msun}{{\ifmmode{{\rm {M_{\odot}}}}\else{${\rm{M_{\odot}}}$}\fi}}  
\newcommand{\lya}{{Lyman $\alpha$~}}

\newcommand{\rand}{{\ifmmode{{\mathcal{R}}}\else{${\mathcal{R}}$ }\fi}}  
\newcommand{\hs}{{\hspace{1mm}}}  
\newcommand{\kms}{{\ifmmode{{\mathrm{\,km\ s}^{-1}}}\else{\,km~s$^{-1}$}\fi}}
\def\lsim{~\rlap{$<$}{\lower 1.0ex\hbox{$\sim$}}}
\def\gsim{~\rlap{$>$}{\lower 1.0ex\hbox{$\sim$}}}
\begin{document}

\title{The impact of gas bulk rotation on the Lyman $\alpha$ line} 
\shorttitle{Rotation effects on the Lyman $\alpha$ line}

\shortauthors{Garavito-Camargo, Forero-Romero \& Dijkstra}

\author{ Juan N. Garavito-Camargo, Jaime E. Forero-Romero}
\affil{Departamento de F\'{i}sica, Universidad de los Andes, Cra. 1
No. 18A-10, Edificio Ip, Bogot\'a, Colombia}
\email{jn.garavito57@uniandes.edu.co}
\email{je.forero@uniandes.edu.co}
\and
\author{Mark Dijkstra}
\affil{Institute of Theoretical Astrophysics, University of Oslo,
  Postboks 1029, 0858 Oslo, Norway} 
\email{mark.dijkstra@astro.uio.no }

\begin{abstract}

We present results of radiative transfer calculations to measure the
impact of gas bulk rotation on the morphology of the Lyman $\alpha$ 
emission line in distant galaxies. 
We model a galaxy as a sphere with an homogeneous mixture of dust and
hydrogen at a constant temperature. 
These spheres undergo solid-body rotation with maximum velocities in
the range $0-300$ \kms and neutral hydrogen optical depths in the
range $\tau_{\rm H}=10^{5}-10^{7}$. 
We consider two types of source distributions in the sphere: central and 
homogeneous.
Our main result is that rotation introduces a dependence of the
line morphology with  viewing angle and rotational velocity. 
Observations with a line of sight parallel to the rotation axis yield
line morphologies similar to the static case. 
For lines of sight perpendicular to the rotation axis both the intensity at the line
center and the line width increase with rotational velocity.   
Along the same line of sight, the line becomes single peaked  at rotational
velocities close to half the line width in the static case. 
Notably, we find that rotation does not induce any spatial anisotropy in the integrated line flux, the escape fraction or the average number of scatterings. This is because Lyman α scattering through a rotating solid-body proceeds identical as in the static case. The only difference is the doppler shift from the different
regions in the sphere that move with respect to the observer. This
allows us to derive an analytic approximation for the viewing-angle
dependence of the emerging spectrum, as a function of rotational velocity.
\end{abstract}

\section{Introduction}
\label{sec:intro}

The detection of strong \ly emission lines has become an essential
method in extra-galactic astronomy to find distant star-forming
galaxies
\citep{PartridgePeebles,Rhoads00,Gawiser2007,Koehler2007,Ouchi08,Yamada2012,Schenker2012,Finkelstein2013}.
The galaxies detected using this method receive the 
name of \ly emitters (LAEs). 
A detailed examination of this galaxy population has diverse
implications for galaxy formation, reionization and the large scale
structure of the Universe.  
Attempts to fully exploit the physical information included in the \ly
line require an understanding of all the physical factors involved in
shaping the  line. 
Due to the resonant nature of this line, these physical factors
notably include temperature, density and bulk velocity field of the
neutral Hydrogen in the emitting galaxy and its surroundings.

A basic understanding of the quantitative behavior of the \ly line
has been reached through analytic studies in the case of a static
configurations, such as uniform slabs
\citep[][]{Adams72,Harrington73,Neufeld90} and uniform spheres
\citep{Dijkstra06}. 
Analytic studies of configurations including some kind of bulk flow
only include the case of a sphere with a Hubble like expansion flow
\citep{LoebRybicki}.  

A more detailed quantitative description of the \ly line has been
reached through Monte Carlo simulations \citep{Auer68,Avery68,Adams72}. 
In the last two decades these studies have become popular due to the
availability of computing power. 
Early into the 21st century, the first
studies focused on homogeneous and static media
\citep{Ahn00,Ahn01,Zheng02}. 
Later on, the effects of clumpy media \citep{Hansen06} and
expanding/contracting shell/spherical geometries started to be studied
\citep{Ahn03,Verhamme06,Dijkstra06}.  For a recent review, we refer the interested reader to \citet{review}.
Similar codes have applied these results to semi-analytic models of
galaxy formation \citep{Orsi12, Garel2012} and results of large
hydrodynamic simulations \citep{CLARA,Forero12,Behrens13}. 
Recently, Monte Carlo codes have also been applied to the results of
high resolution hydrodynamic simulations of individual galaxies
\citep{Laursen09,Barnes11,Verhamme12,Yajima12}. 
Meanwhile, recent  developments have been focused on the systematic
study of clumpy outflows \citep{DijkstraKramer} and anisotropic
velocity configurations \citep{Zheng2013}. 

The recent studies of galaxies in hydrodynamic simulations
\citep{Laursen09,Barnes11,Verhamme12,Yajima12} have all shown
systematic variations in the \ly line with the viewing angle. These
variations are a complex superposition of anisotropic density
configurations (i.e. edge-on vs. face-on view of a galaxy), the
inflows observed by gas cooling and the outflows included in the
supernova feedback process of the simulation. These bulk flows
physically correspond to the circumgalactic and intergalactic medium
(CGM and IGM). These effects are starting to be studied
 in simplified configurations that vary the density and wind
 characteristics \citep{Zheng2013,Behrens2014}. 

However, in all these efforts the effect of rotation,
which is an ubiquitous feature in galaxies, has not been
systematically studied. The processing of the \ly photons in a
rotating interstellar medium (ISM) must have some kind of impact in
the \ly line morphology. 

Performing that study is the main goal of this paper. We investigate for the
first time the impact of rotation on the morphology of the \ly
line. We focus on a simplified system: a spherical gas cloud with
homogeneous density and solid body rotation, to study the line
morphology and the escape fraction in the presence of dust. We base
our work on two independent Monte Carlo based radiative transfer codes
presented in \cite{CLARA} and \cite{DijkstraKramer}.   
  
This paper is structured as follows: In \S \ref{sec:implementation} we
present the implementation of bulk rotation into the Monte Carlo
codes, paying special attention to coordinate definitions. We also
present a short review of how the \ly radiative transfer codes work
and list the different physical parameters in the simulated grid of
models. In \S \ref{sec:results} we present the results of the
simulations, with special detail to quantities that show a
clear evolution as a function of the sphere rotational velocity. In \S
\ref{sec:discussion} we discuss the implications of our results. In
the last section we present our conclusions. The Appendix presents the
derivation of an analytic expression to interpret
the main trends observed in the Monte Carlo simulations.

In this paper we express a photon's frequency in terms of the
dimensionless variable $x\equiv (\nu -\nu_a)/\Delta\nu_{\rm D}$, where
$\nu_{\rm \alpha}=2.46\times 10^{15}$ Hz is the Ly$\alpha$ resonance
frequency,  $\Delta\nu_{\rm D} \equiv
\nu_{\alpha}\sqrt{2kT/m_pc^2}\equiv \nu_av_{\rm th}/c $ is the Doppler
broadening of the line which depends on the neutral gas temperature
$T$ or equivalently the thermal velocity
$v_{\rm th}$ of the atoms. We also use the parameter $a$ to define the
relative line width as $a=\Delta\nu_{\alpha}/2\Delta\nu_{\rm D}$,
where $\Delta\nu_{\alpha}$ is the intrinsic linewidth. For the
temperature $T=10^4$K used in our radiative transfer calculations the
thermal velocity is $v_{\rm th}=12.8$\kms.

\section{Models of bulk gas rotation}
\label{sec:implementation}

Describing the kinematics of gas rotation in all generality is a
complex task, specially at high redshift where there is still missing
a thorough observational account of rotation in galaxies beyond
$z>1.0$. Even at low redshifts there is a great
variation in the shape of the rotation curve as observed in HI
emission as a function of the distance to the galaxy center. However
there are two recurrent features. First, in the
central galactic region the velocity increases proportional to the radius,
following a solid rotation behavior. Second, beyond a certain radius
the rotation curve tends to flatten.  An ab-initio description of
such realistic rotation curves in simulations depends on having access to
the dynamic evolution of all mass components in the galaxy: stars, gas
and dark matter. Such level of realism is extremely complex to
achieve, specially if one wants to get a systematic description based
on statistics of simulated objects. 

Following the tradition of studies of \ly emitting systems,
we implement a model with simplified geometry. We assume that the gas
is homogeneously distributed in a sphere that rotates as a solid body
with constant angular velocity. This simple model will contain only
one free parameter: the linear velocity at the sphere's surface, $V_{\rm
  max}$. 

\subsection{Detailed Implementation of Rotation}

 In the Monte Carlo code we define a Cartesian coordinate system to
 describe the position of each photon. The origin of this system
 coincides with the center of the sphere and the rotation axis is defined
 to be $z$-axis. With this choice, the components of the gas bulk velocity
 field, $\vec{v} = v_{x}\hat{i} + v_{y}\hat{j} + v_{z}\hat{k}$, can be
 written as  
  
\begin{equation}
    v_{x}=-\frac{y}{R}V_{\rm max}, \label{subeq1}
\end{equation}
\begin{equation}
    v_{y}=\frac{x}{R}V_{\rm max}, \label{subeq2}
\end{equation}
\begin{equation}
    v_{z}=0, \label{subeq3}
\end{equation}
where $R$ is the radius of the sphere and $V_{\rm max}$ is the linear
velocity at the sphere's surface. The minus/plus sign in the
$x$/$y$-component of the velocity indicates the direction of
rotation. In this case we take the angular velocity in the same
direction as the $\hat{k}$ unit vector. With these definitions we can
write the norm of the angular velocity as $\omega=V_{\rm max}/R$.  

\begin{figure}
\begin{center}
  \includegraphics[width=0.4\textwidth]{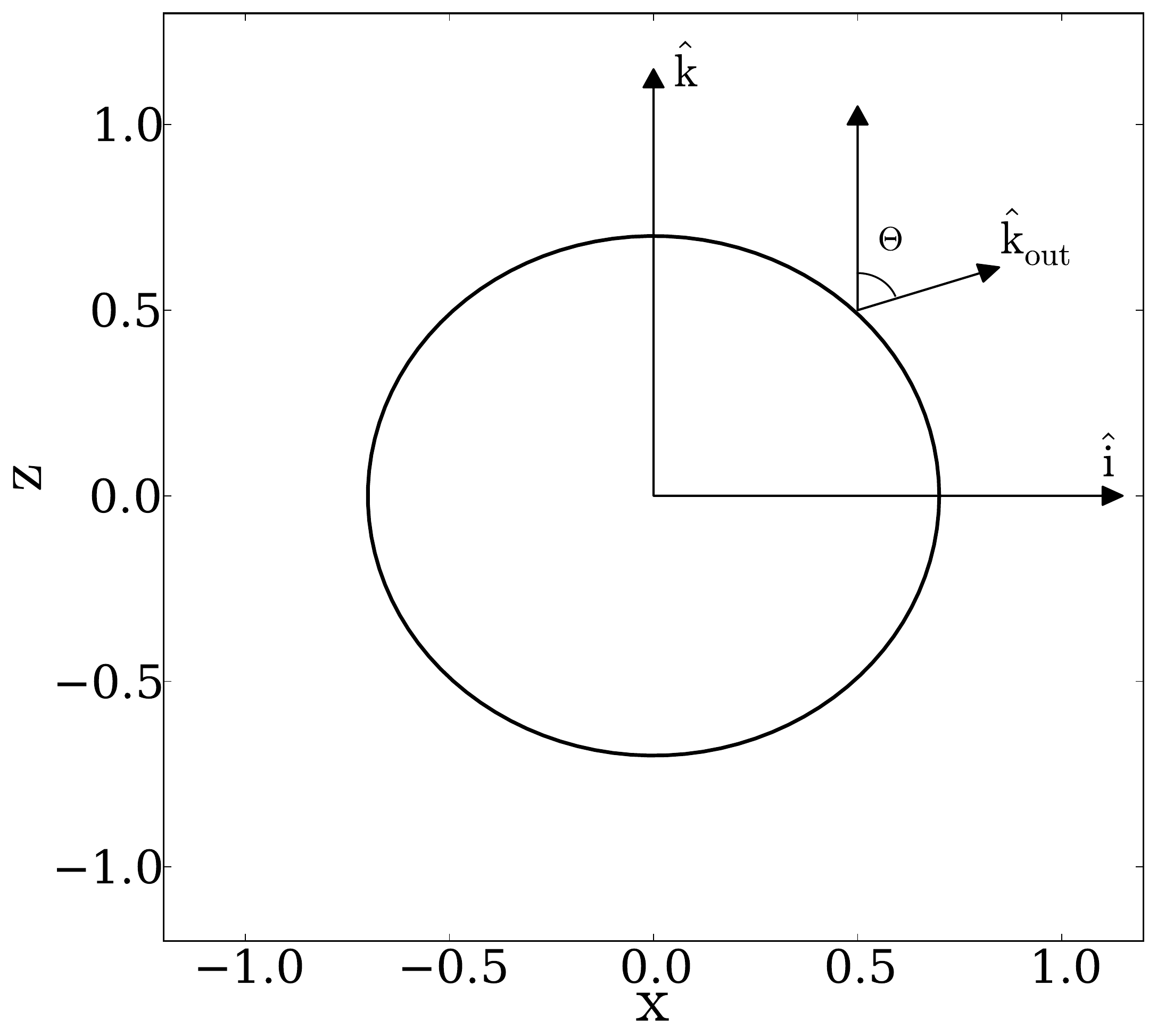}
\end{center}
\caption{Geometry of the gas distribution. The angular velocity vector
  is parallel to the unit vector $\hat{k}$. In order to describe the
  departures from spherical symmetry we use the polar angle $\theta$
  formed by the direction of the outgoing photons with respect to the
  $z$-axis. We define define the variable $\mu\equiv\cos\theta$ to
  report to present our results. Computing the spectra for photons in
  a   narrow range of $\mu$ is equivalent to having a line-of-sight
  oriented in that direction.  
    \label{fig:geometry}}  
\end{figure}

For each photon in the simulation we have its initial position inside
the sphere, direction of propagation $\hat{k}_{\rm in}$ and reduced
frequency $x_{\rm in}$. 
The photon's propagation stops once they cross the
surface of the sphere. At this point we store the position, the outgoing direction
of propagation $\hat{k}_{\rm out}$ and the reduced frequency $x_{\rm
  out}$. We now define the angle $\theta$ by $\cos\theta = \hat{k}_{\rm out}\cdot
\hat{k}\equiv \mu$, it is the angle of the outgoing photons with 
respect to the direction of the angular velocity. We use the variable $\mu$ to
study the anisotropy induced by rotation. Fig. \ref{fig:geometry}
shows the geometry of the problem and the important variables.

\begin{figure*}
\begin{center}
  \includegraphics[width=0.95\textwidth]{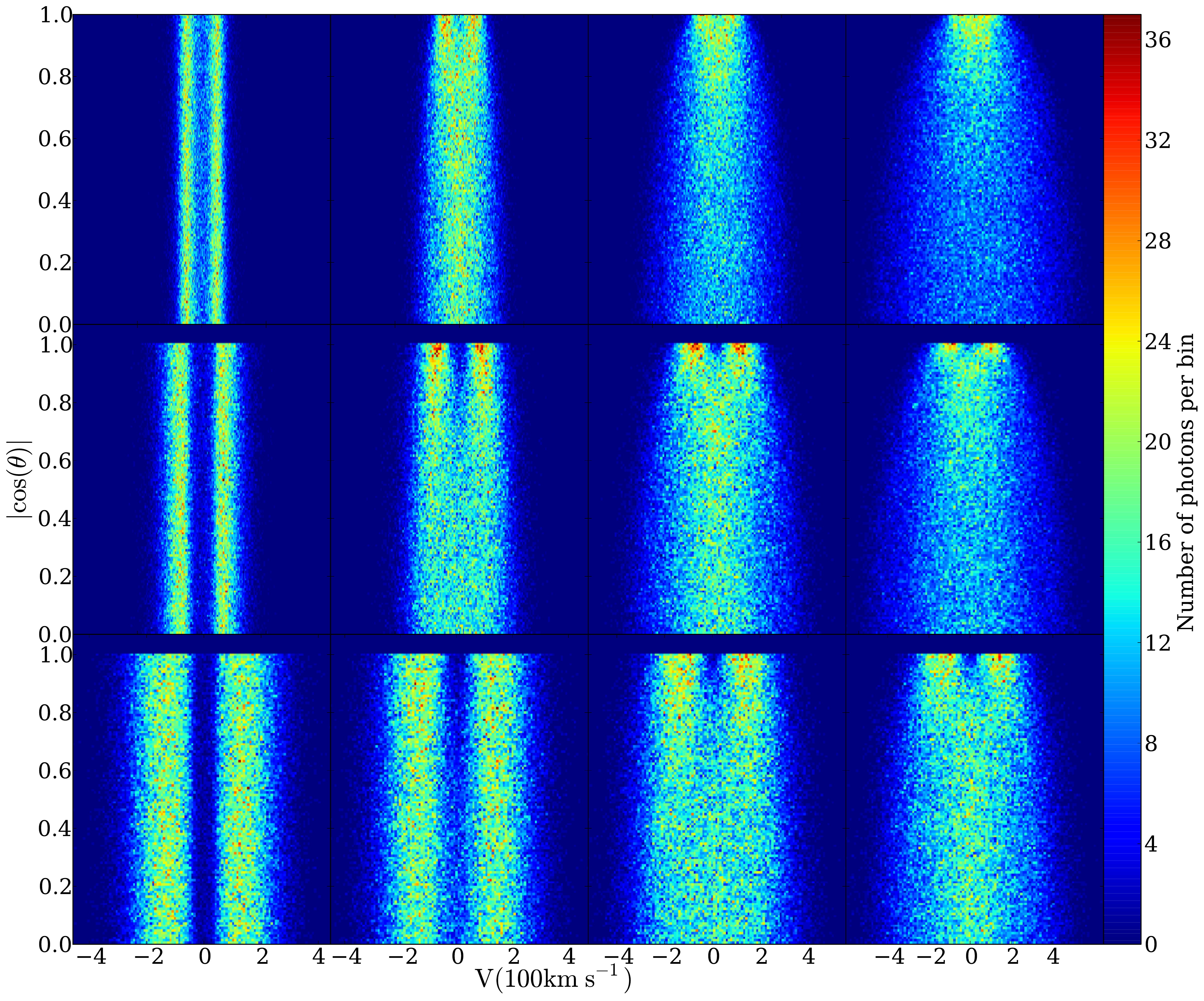}
\end{center}
\caption{
2D histogram showing the number of photons that escape with frequency
$x$ forming an angle $\theta$ (parametrized as $|\cos\theta|$) with the
rotation axis. 
The rotational velocity ($0,100,200,300$\kms) increases from left to
right and the optical depth ($10^5$, $10^6$, $10^7$) from top to
bottom. 
The \ly photons are initialized at the center of the sphere. 
Two main results can be read from this figure.
First, the line morphology depends on the viewing angle. 
Second, the line can become single peaked for high rotational
velocities.
\label{fig:CentralSpec} }   
\end{figure*}

\begin{figure*}
\begin{center}
  \includegraphics[width=0.95\textwidth]{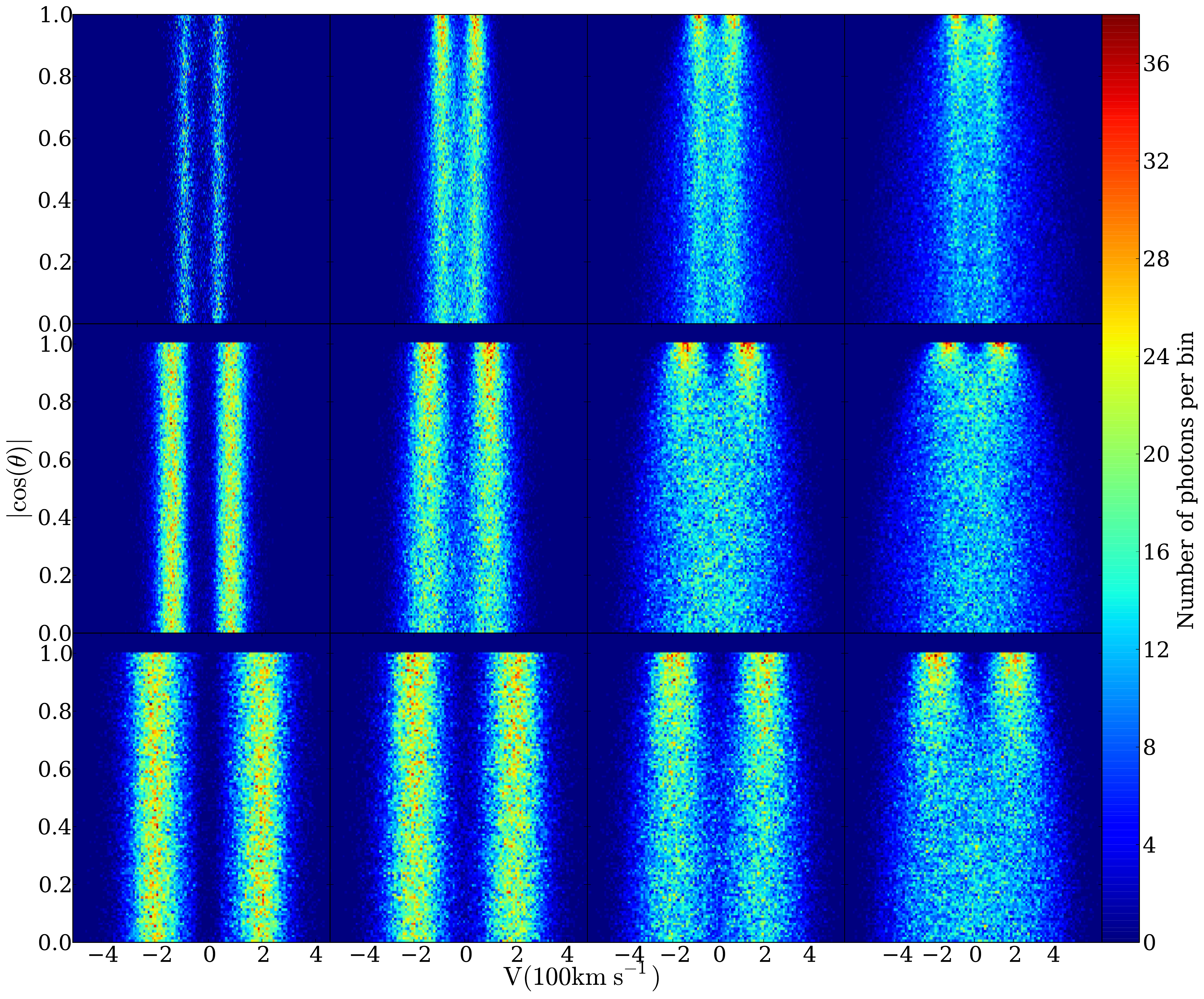}
\end{center}
\caption{Same as Fig. \ref{fig:CentralSpec} for \ly photons
  initialized homogeneously throughout the sphere.
    \label{fig:HomSpec}}  
\end{figure*}

\begin{figure*}
\begin{center}
  \includegraphics[width=0.95\textwidth]{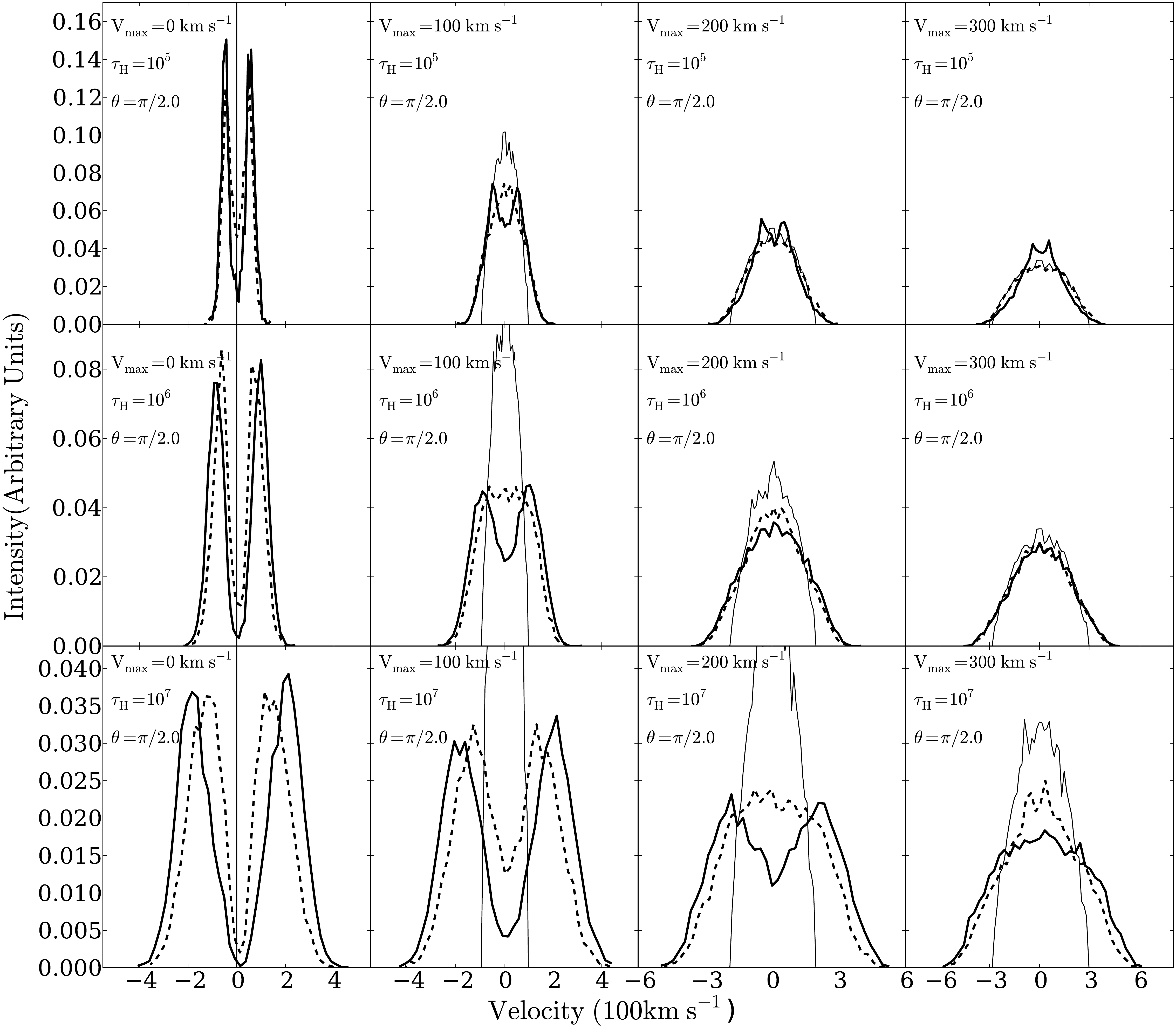}
\end{center}
\caption{Shape of the \ly line for different maximum rotational
  velocities for a LoS perpendicular to the rotation axis
  ($|\mu|\sim 0$). The continuous (dashed) line represents the central
  (homogeneous) source distributions. The continuous thin line
represents the intrinsic homogeneous spectrum. The panels follow the same
  distribution as in Fig.s \ref{fig:CentralSpec} and \ref{fig:HomSpec}.
    \label{fig:differentvelocities}}  
\end{figure*}

\begin{figure*}
\begin{center}
  \includegraphics[width=0.95\textwidth]{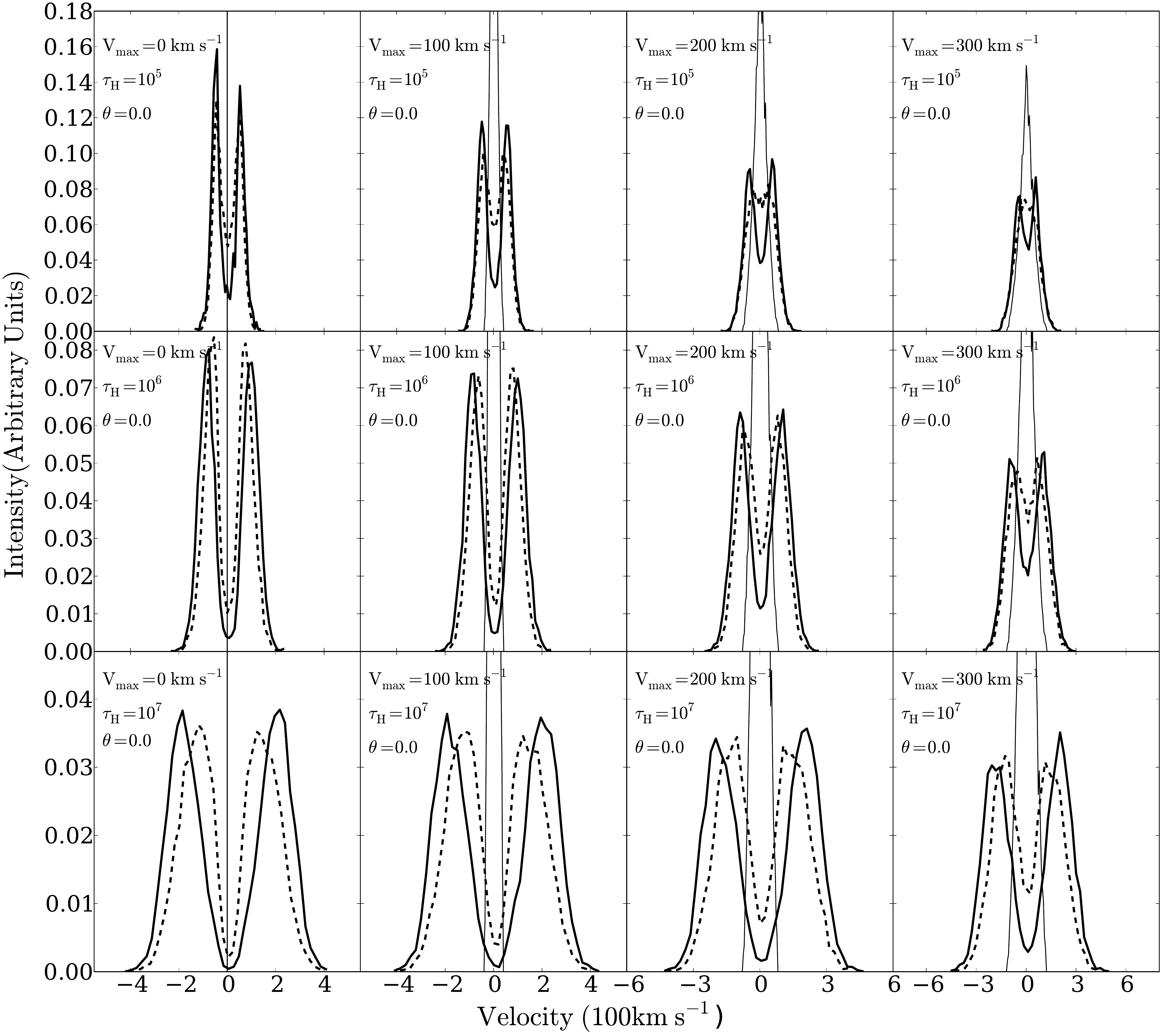}
\end{center}
\caption{Shape of the \ly line for different maximum rotational
  velocities for a LoS perpendicular to the rotation axis
  ($|\mu|\sim 1$). The continuous (dashed) line represents the central
  (homogeneous) source distributions.  The continuous thin line
represent the intrinsic homogeneous spectrum. The panels follow the same
  distribution as in Fig.s \ref{fig:CentralSpec} and \ref{fig:HomSpec}.
    \label{fig:differentvelocities2}}
\end{figure*}

\subsection{Brief Description of the Radiative Transfer Codes}

Here we briefly describe the relevant characteristics of the two
radiative transfer codes we have used.  
For a detailed description we refer the reader to the original papers
\cite{CLARA} and \cite{DijkstraKramer}. 

The codes follow the individual scatterings of \ly photons as they
travel through a 3D distribution of neutral Hydrogen. 
The frequency of the photon (in the laboratory frame) and
its direction of propagation change at every scattering. 
This change in frequency is due to the peculiar velocities of the
Hydrogen absorbing and re-emitting the photon. 
Once the photons escape the gas distribution we store their direction of
propagation and frequency at their last scattering.

The initialization process for the \ly photons specifies its
position, frequency and direction of propagation. 
We select the initial frequency to be exactly the \ly rest-frame
frequency in the gas reference frame and the direction of propagation
to be random following an flat probability distribution over the sphere.  
It means that for photons emitted from the center of the sphere
$x_{\rm in}=0$, while photons emitted at some radii with a peculiar
velocity $\vec{v}$ have initial values $x_{\rm in}$ depending on its direction of
propagation: $x_{\rm   in}=\vec{v}\cdot\hat{k}_{\rm in}/v_{\rm th}$.
We do not include the effect of turbulent velocities in the
initialization.
We neglect this given that the induced perturbation should be on the
close to the thermal velocity, $12.8$\kms, which is one order of
magnitude smaller than the velocity widths ($100$-$500$\kms) in the
static case.

If dust is present, the photon can interact either with a Hydrogen
atom or dust grain.
In the case of a dust interaction the photon can be either absorbed or
scattered.
This probability is encoded in the dust albedo, $A$, which we chose
to be $1/2$. 
In order to obtain accurate values for the escape fraction of
photons in the presence of dust, we do not use any accelerating
mechanism in the radiative transfer.

The codes treat the gas as homogeneous in density and temperature.  
This implies that the gas is completely defined by its geometry
(i.e. sphere or slab), temperature $T$, Hydrogen optical depth
$\tau_{\rm H}$, dust optical depth $\tau_{\rm a}$ and the bulk
velocity field $\vec{v}$.

\subsection{Grid of Simulated Galaxies}
\label{sec:models}

In the Monte Carlo calculations we follow the propagation of $N_{\gamma}=10^5$
numerical photons through different spherical galaxies. 
For each galaxy we vary at least one of the following parameters: the maximum
rotational velocity $V_{\rm max}$, the hydrogen optical depth $\tau_{H}$,
the dust optical depth $\tau_{a}$ and the initial distribution of photons
with respect to the gas. 
In total there are $48$ different models combining all the possible
different variations in the input parameters.
Table \ref{table:models} lists the different parameters we used to
generate the models. The results and trends we report are observed in both 
Monte Carlo codes.

\begin{table}
\begin{center}
\begin{tabular}{llc}\hline\hline
Physical Parameter (units) & Symbol & Values\\\hline
Velocity (\kms) & $V_{\rm max}$&$0,\ \ 100,\ 200,\ 300$\\
Hydrogen Optical Depth & $\tau_{H} $ & $10^{5},\ 10^{6},\ 10^{7}$\\
Dust Optical Depth & $\tau_{a}$ & $0$,$1$\\
Photons Distributions & & Central, Homogeneous\\\hline\hline
\end{tabular}
\caption{
  Summary of Physical Parameters of our Monte Carlo Simulations.} 
\label{table:models}
\end{center}
\end{table}

\section{Results}
\label{sec:results}

The main results of this paper are summarized in Fig.
\ref{fig:CentralSpec} and \ref{fig:HomSpec}. 
They show 2D histograms of the escape frequency $x$ and outgoing angle
$\theta$ parametrized by $|\mu|$. 
Taking into account only photons photons around a value
of $|\mu|$ gives us the emission detected by an observer located at an
angle $\theta$ with respect to the rotation axis. 
We have verified that the solutions are indeed symmetric with respect
to $\mu=0$. We have also verified that the total flux is the same for all $\mu$.

From these figures we can see that the line properties change with
rotational velocity and depend on the viewing angle $\theta$.  
In the next subsections we quantify the morphology changes with with
velocity, optical depth and viewing angle.  
We characterize the line morphology by its total intensity, the full
width at half maximum, (FWHM) and the location of the peak maxima. 
In order to interpret the
morphological changes in the line we also report the median number of
scatter for each \ly photon in the simulation. 
For the models where dust is included we measure the escape fraction
as a function of rotational velocity and viewing angle.   

\subsection{Line Morphology}
\label{sec:angles}

The first column in both Fig. \ref{fig:CentralSpec} and
\ref{fig:HomSpec} shows that for the static sphere the line properties
are independent of $|\mu|$, as it is expected due
to the spherical symmetry. 
However, for increasing rotational velocities, at a fixed optical
depth, there are clear signs that this symmetry is broken. 

If the viewing angle is aligned with the rotation axis, $|\mu|\sim
1$, the \ly line keeps a double peak with minor
changes in the morphology as the rotational velocity increases. 
However, for a line of sight perpendicular to the rotation axis,
$|\mu|\sim 0$, the impact of  rotation is larger.  
The double peak readily transforms into a single peak.

This is clear in  Fig. \ref{fig:differentvelocities} where we
present the different line morphologies for $|\mu|\sim 0$ in the
homogeneous and central configurations.  
The panels have the same distribution as Fig. \ref{fig:CentralSpec}
and \ref{fig:HomSpec}.    
There are three clear effects on the line morphology as the rotational
velocity increases.  
First, the line broadens; second, the double peaks reduce their intensity; and
third, the intensity at the line centre rises. 
The last two effects are combined to give the impression that the double
peaks are merged into a single one at high rotational velocities.

\subsection{Integrated Line Intensity}
\label{sec:intlineint}

\begin{figure}
\begin{center}
  \includegraphics[width=0.4\textwidth]{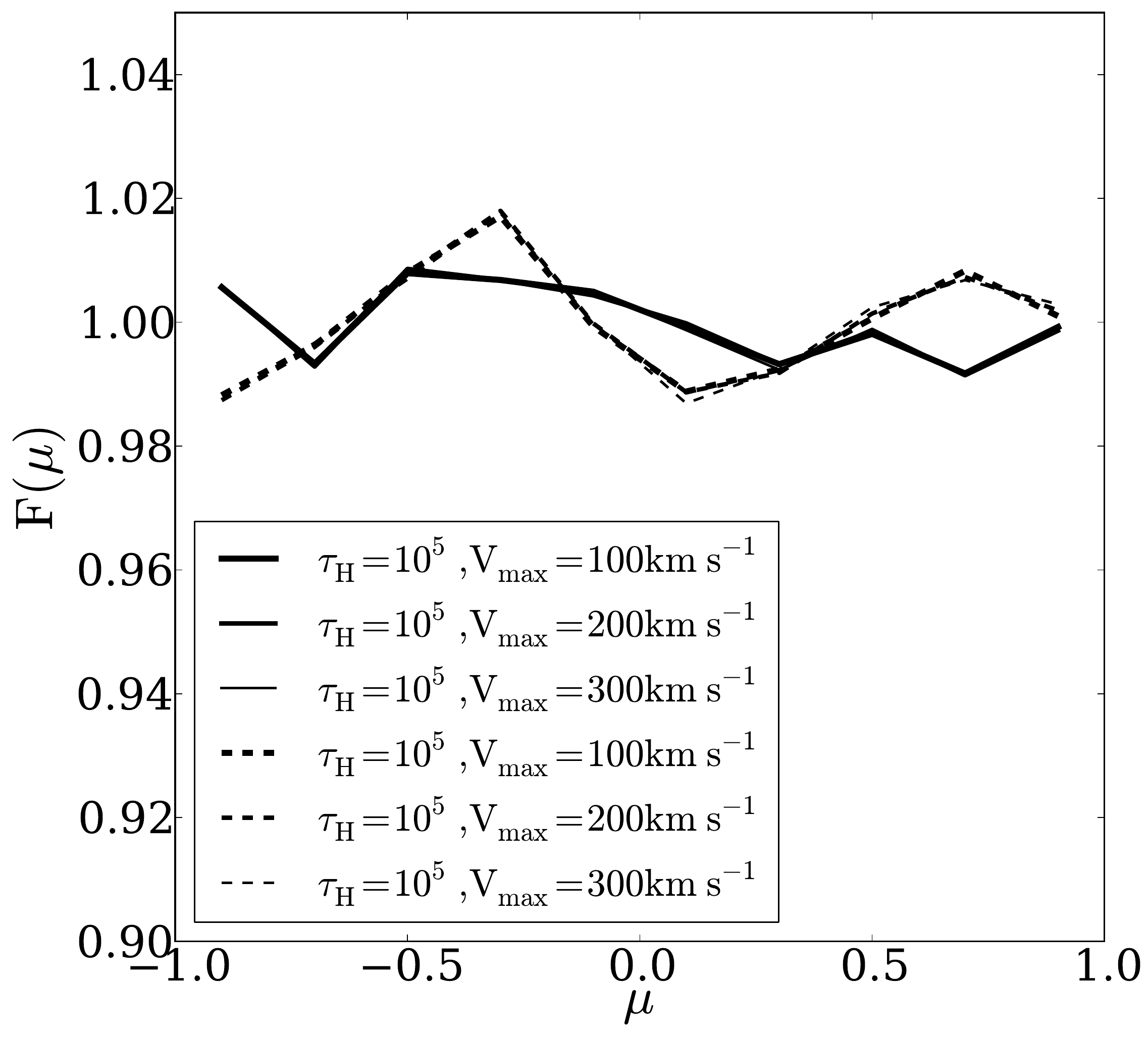}
\end{center}
\caption{Integrated flux distribution as a function of the
  viewing angle as parametrized by $\mu$. Continuous (dashed)
  correspond to central (homogeneous) source distribution.
  The models correspond to an optical depth of $\tau_{\rm
    H}=10^5$ and rotational velocities of $100$\kms, $200$\kms and
  $300$\kms. The distributions are flat in the range of models probed
  in this paper, meaning that the integrated flux for all viewing
  angles is the same.
\label{fig:muhisto}} 
\end{figure}

We now consider possible variations in the integrated flux with
respect to the viewing angle $\theta$. 
To this end we  define the normalized flux seen by an observer at an
angle $\mu$ by:

\begin{equation}
F(\mu) = \frac{2\Delta N}{N\Delta \mu}, 
\end{equation} 
where $\mu=\cos\theta$, $N$ is the total number of outgoing photons,
$\Delta N$ is the number of photons in an angular bin $\Delta
\theta$. This definition satisfies the condition
$\int_{-1}^{1}F(\mu)d\mu/2=1$.  In the case of perfect spherical
symmetry one expects a flat distribution with $F(\mu)=1$.

Fig. \ref{fig:muhisto} shows the results for a selection of models
with $\tau_{\rm H}=10^{5}$, different rotational velocities and the two
types of source distributions. This shows that $F(\mu)$ is consistent with being flat, apart
from some statistical fluctuations on the order of 2\%. 

This is a remarkable result: while the rotation axis defines preferential direction, the
integrated flux is the same for all viewing angles in the range of parameters explored in this paper. This can be understood from the fact that
{\it radiative transfer inside a sphere that undergoes solid-body
  rotation proceeds identical as inside a static sphere}: we can draw
a line between any two atoms within the rotating cloud, and their
relative velocity along this line is zero (apart from the relative
velocity as a result of random thermal motion), irrespective of the
rotation velocity of the cloud. This relative velocity is what is
relevant for the radiative transfer\footnote{This point can be further
  illustrated by considering the path of individual photons: let a
  photon be emitted at line center ($x=0$), in some random direction
  ${\bf k}$, propagate a distance that corresponds to $\tau_0=1$,
  scatter fully coherently (i.e. $x=0$ after scattering in the gas
  frame) by 90$^{\circ}$, and again propagate a distance that
  corresponds to $\tau_0=1$. The position where the photon scatters
  next does {\it not} depend on the rotation of the cloud, nor on
  ${\bf k}$.}
	
\subsection{Full Width at Half Maximum}
\label{sec:widthpeak}

\begin{figure*}
\begin{center}
  \includegraphics[width=0.95\textwidth]{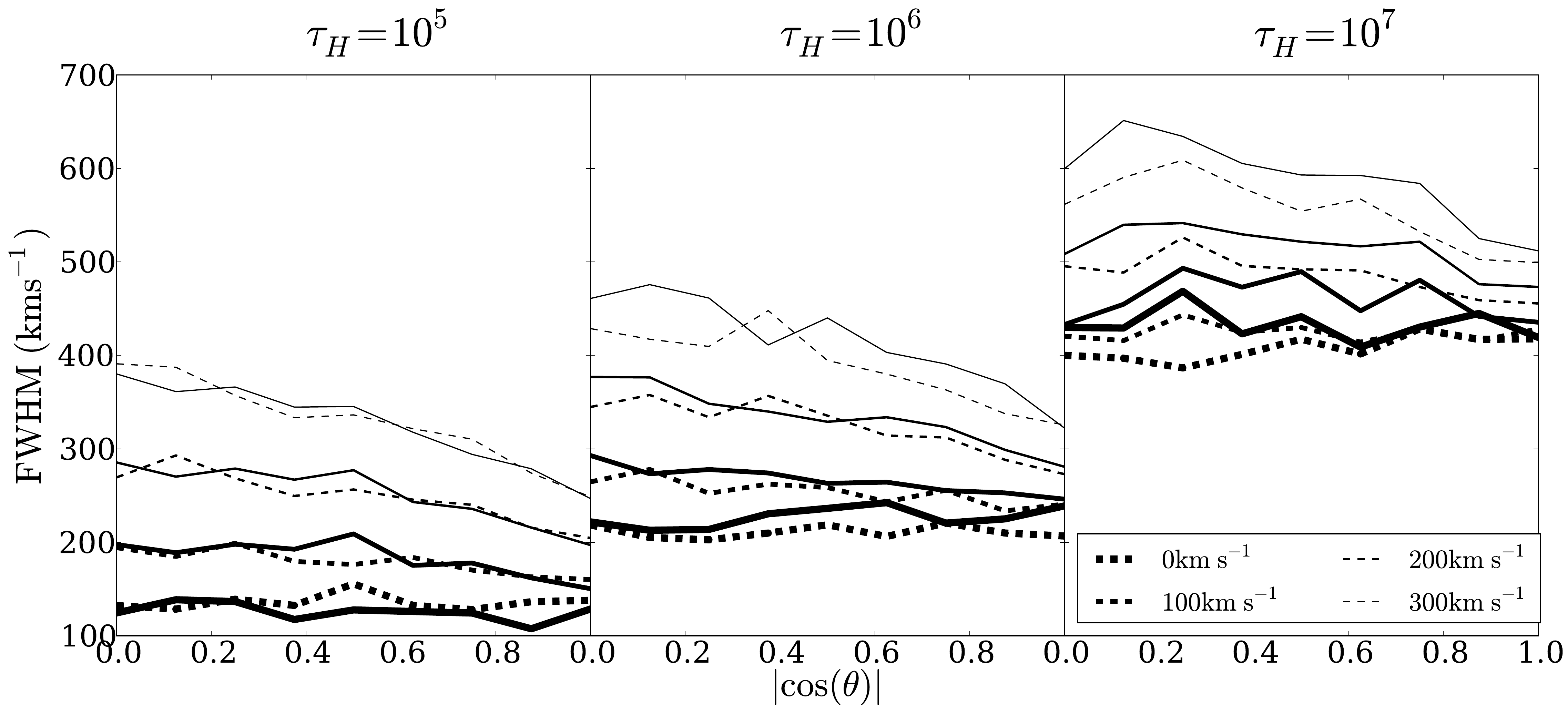}
\end{center}
  \caption{FWHM for the non-dusty models as a function of the viewing
  angle parametrized by $|\cos\theta|$. Continuous (dashed) lines  correspond
  to central (homogeneous) source distributions. The general trend is
  of an decreasing line width as the line of sight becomes parallel to the
  rotation axis.
  \label{fig:widthvsmu}} 
\end{figure*}

\begin{figure*}
\begin{center}
  \includegraphics[width=0.95\textwidth]{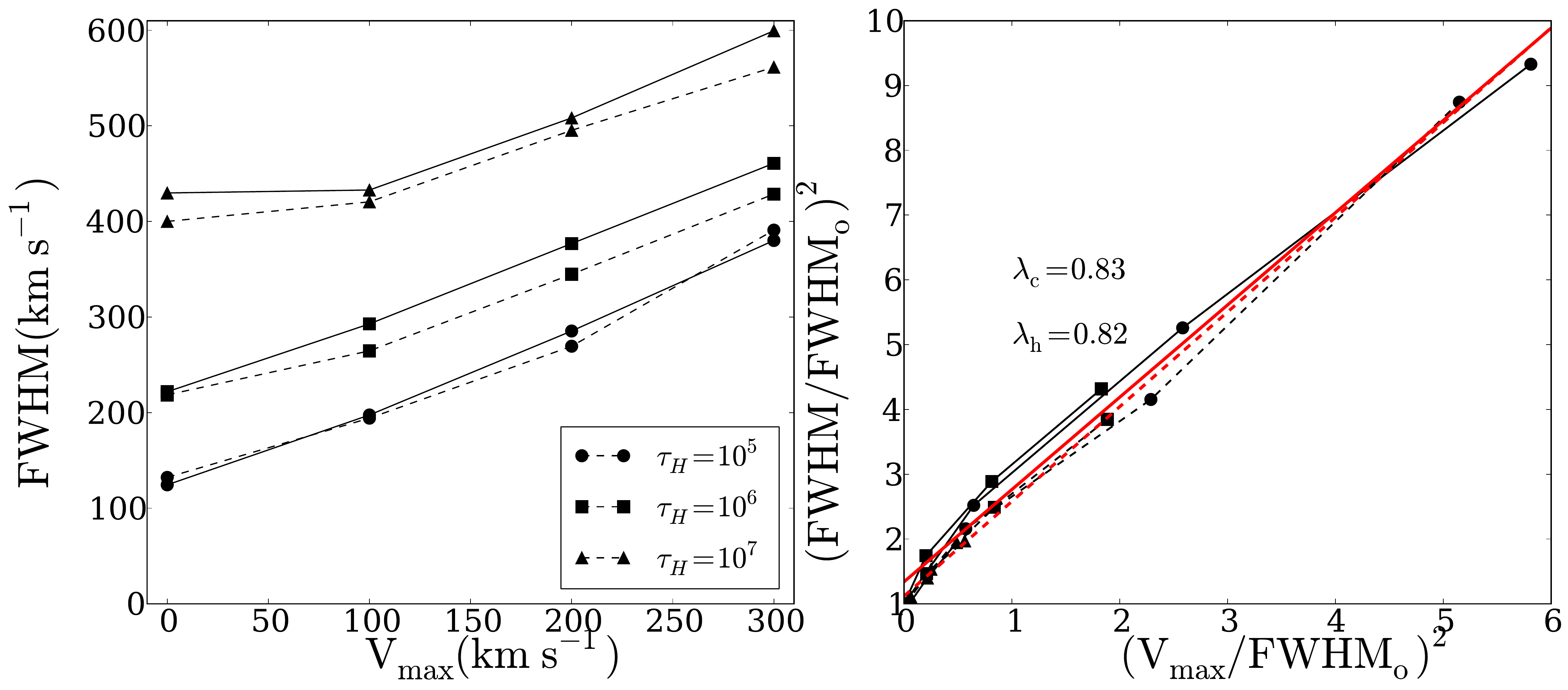}
\end{center}
\caption{FWHM for the non-dusty models as a function of
  rotational velocity $V_{\rm max}$ for observers located
  perpendicular to the rotation axis. 
  The left panel shows the results in velocity units while the right
  panel normalizes the data by the FWHM in the static case. 
  Continuous (dashed) lines  correspond to central (homogeneous)
  source distributions. 
  The straight lines represent the fit to the data using the
  expression in Eq. (\ref{eq:fwhm}). 
  \label{fig:widthsvsvelocity}}
\end{figure*}

We use the full width at half maximum (FWHM) to quantify the line
broadening. 
We measure this width from the line intensity histogram by finding the
values of the velocities at half maximum intensity.
We use lineal interpolation between histogram points to get a value
more precise than the bin size used to construct the histogram. 

Fig. \ref{fig:widthvsmu} shows the FWHM for all models as a function
of the viewing angle. 
The FWHM increases for decreasing values of $\mu$ (movement from the
poles to the equator) and increasing values of $V_{\rm max}$. 
In Fig.
\ref{fig:widthsvsvelocity} we fix $|\mu|<0.1$, i.e. viewing angle
perpendicular to the rotation axis, to plot the FWHM as a function of
rotational velocity.

We parametrize the dependency of the line width with  $V_{\rm max}$ as
\begin{equation}
 {\rm FWHM}^2 = {\rm FWHM}_{0}^2 + V_{\rm max}^2/\lambda^2,
\label{eq:fwhm}
\end{equation}
where FWHM$_{0}$ is the velocity width in the static case and $\lambda$ 
is a positive scalar to be determined as a fit to the data.  
With this test we want to know to what extent the new velocity width can be
expressed as a quadratic sum of the two relevant velocities in the
problem. 

All the models fall into a single family of lines in the plane shown
in the right panel of Fig. \ref{fig:widthsvsvelocity}, justifying
the choice of our parametrization.
We fit simultaneously all the points in two separate groups, central
and homogeneous sources.
We find that these values are $\lambda_{\rm c} = 0.83 \pm 0.06$ and
$\lambda_{\rm    h}= 0.82\pm 0.05$ respectively.

\subsection{Line Maxima}
\label{sec:maxima}

\begin{figure*}
\begin{center}
  \includegraphics[width=0.95\textwidth]{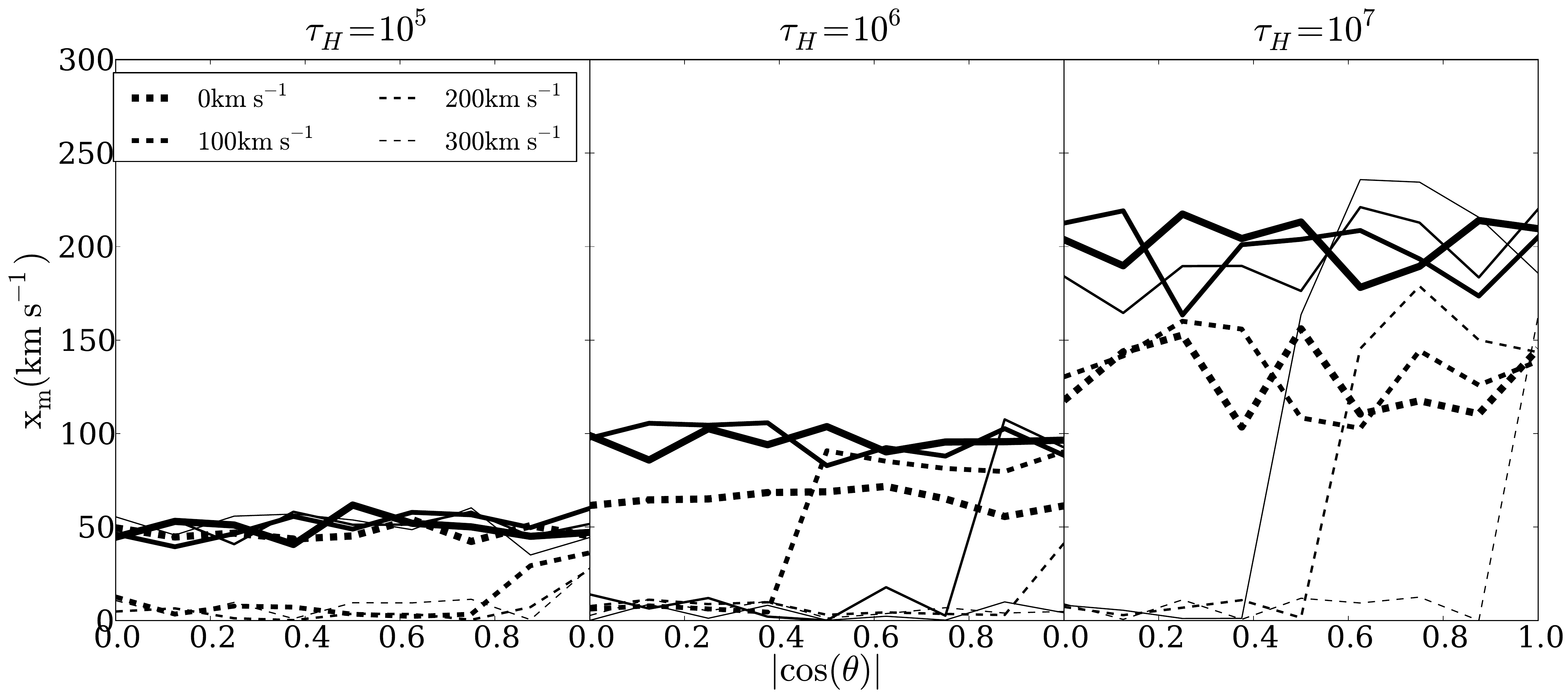}
\end{center}
\caption{Position of the line maxima as a function of maximum
  rotational   velocity $V_{\rm max}$. Continuous (dashed) lines
  correspond to   central (homogeneous) source distributions. A value
  of $x_{\rm     max}=0$ indicates that line becomes single
  peaked. \label{fig:maximumvsvelocity}}
\end{figure*}

We measure the peak maxima position, $x_m$, to quantify the transition from
double into single peak profiles. 
In Fig. \ref{fig:maximumvsvelocity} we show the dependence of $x_m$ with
the viewing angle parametrized by $|\cos\theta|$ for different
rotational velocities.
There are two interesting features that deserve attention. 
First, for a viewing angle parallel to the rotational axis ($\mu\sim
1.0$) the maxima of all models with the same kind of source
initialization are similar regardless of the rotational velocity.  
Second, at a viewing angle perpendicular to the rotation axis ($\mu\sim 0.0$) a
large fraction of models become single peaked. 
This feature appears more frequently for homogeneously distributed
sources if all the other parameters are equal. 

\subsection{Dusty Clouds: Escape Fraction}
\label{sec:escapefraction}

\begin{table}
\begin{center}
\begin{tabular}{c cccccc}
\hline \hline
Source & $\tau_{H}$ & &  $\ V_{\rm max}$& & \\
Distribution& &    & (\kms) & & \\ 
& & 0 & 100 &200 & 300\\ \hline 
Homogeneous & $10^{5}$& 0.263 &  0.263 &  0.263 &  0.263  \\
            & $10^{6}$ & 0.291 &   0.292 &  0.293 &  0.293 \\
            &$10^{7}$ &  0.228 &  0.228 &  0.228 &  0.228 \\
Central & $10^{5}$ &  0.096 & 0.096 &  0.096 & 0.096 \\
  		&$10^{6}$ & 0.066 &  0.066 &  0.066 &  0.066 \\
 		&$10^{7}$ & 0.015 & 0.016 & 0.016 & 0.015 \\
\hline
\end{tabular}
\caption{
 Escape fraction values for all dusty models. } 
\label{table:escape}
\end{center}
\end{table}

We now estimate the escape fraction $f_{\rm esc}$ for the dusty
models. The main result is that we do not find any significant dependence
with either the viewing angle nor the rotational velocity. This is consistent with our finding in \S~\ref{sec:intlineint}, that radiative transfer inside the cloud does not depend on its rotational velocity. For completeness we list in Table \ref{table:escape} the escape
fraction for all models.

We now put these results in the context of the analytic solution for
the infinite slab\citep{Neufeld90}.   
In Neufeld's set-up the analytic solution depends
uniquely on the product $(a\tau_{\rm   H})^{1/3}\tau_{A}$ where
$\tau_{A} = (1 - A)\tau_{a}$, valid only in the limit $a\tau_{\rm
  H}\gg 1$. 
At fixed values of $\tau_{a}$ the escape fraction monotonically
decreases with increasing values of $\tau_{\rm H}$. 
This expectation holds for the central sources. 
But in the case of homogeneous sources the escape fraction increases
slightly from $\tau_{\rm H}=10^5$ to $\tau_{\rm H}=10^{6}$ 

The naive interpretation of the analytic solution does not seem to
hold for photons emitted far from the sphere's center. 
We suggest that increasing $\tau_{\rm H}$ from $10^{5}$ to $10^{6}$ causes a
transition from the 'optically thick' to  the 'extremely optically
thick' regime for a noticeable fraction of the photons in the
homogeneous source distribution.

In the optically thick regime, \ly  photons can escape in
'single flight' which corresponds to a scenario in which the
photon resonantly scatters $10^4-10^5$ times until it is scattered
into the wing of the line ($x\sim 3-4$). 
At these frequencies the medium is optically thin, and the photons can
escape efficiently in a single flight. 
In contrast, in an extremely optically thick medium \ly
  photons escape in a `single excursion' \citep{Adams72}. 
Here, photons that are scattered into the wing of the line escape from
the medium in a sequence of wing scattering events. 
In both cases, \ly photons resonantly scatter $10^4$-$10^5$ times. 
Because we keep our clouds the same size, the mean free path of Lya
photons that scatter resonantly is 10 times larger for the case
$\tau_{\rm H}=10^5$ than for $\tau_{\rm H}=10^6$. 
If we compute the average distance $D$ travelled by \ly
photons through a medium of size $R$ as a function of line center
optical depth $\tau_{\rm H}$, then we find that during the transition
from optically thick to extremely optically thick the mean traversed
distance $D$ actually decreases slightly.  
This decrease is unique to this transition region, and $D$ generally
increases with $\tau_{\rm H}$ at other values of $\tau_{\rm H}$.   

\subsection{Average Number of Scatterings}
\label{sec:scatterings}

\begin{figure*}
\begin{center}
    \includegraphics[width=0.45\textwidth]{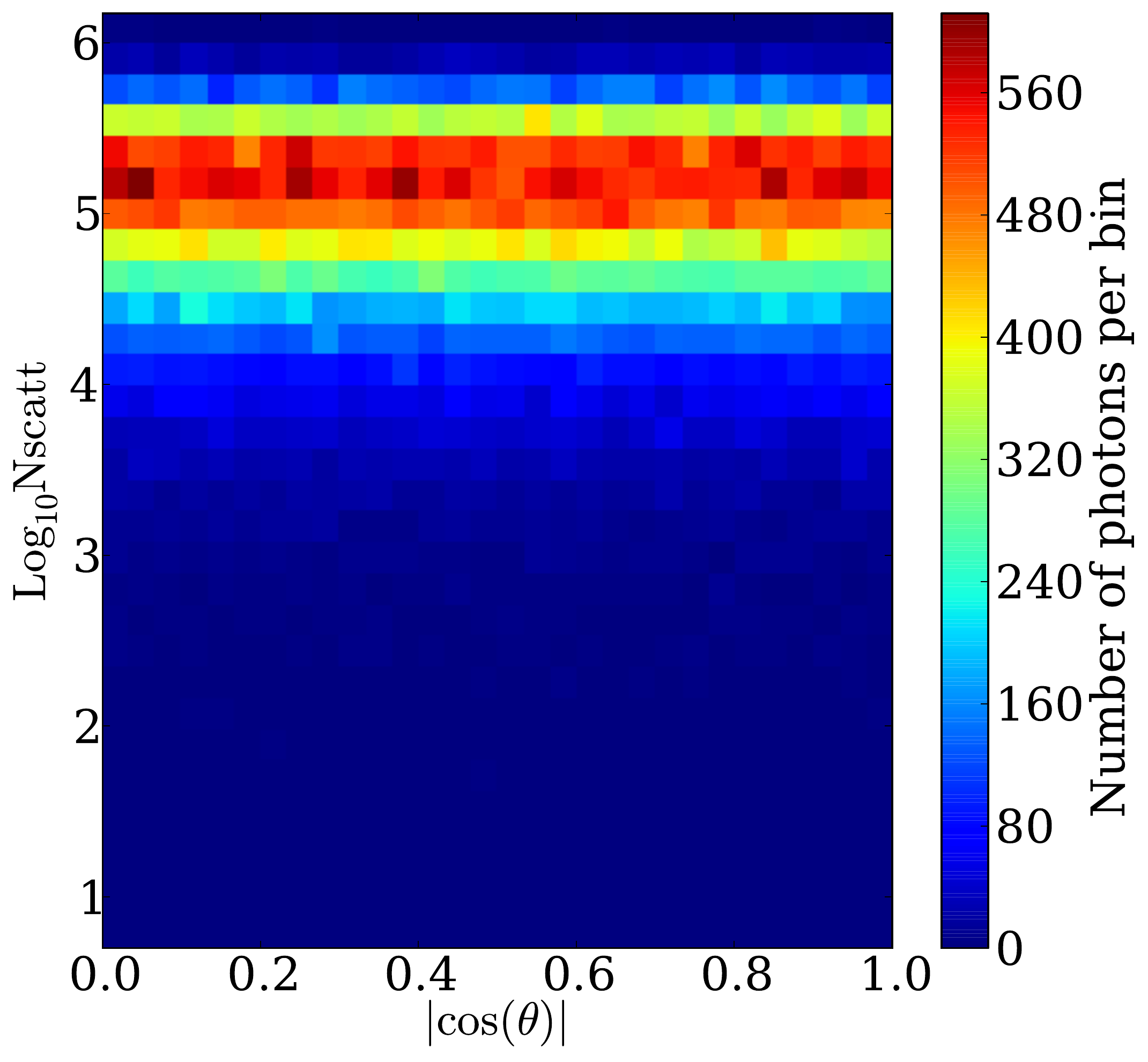}
    \includegraphics[width=0.45\textwidth]{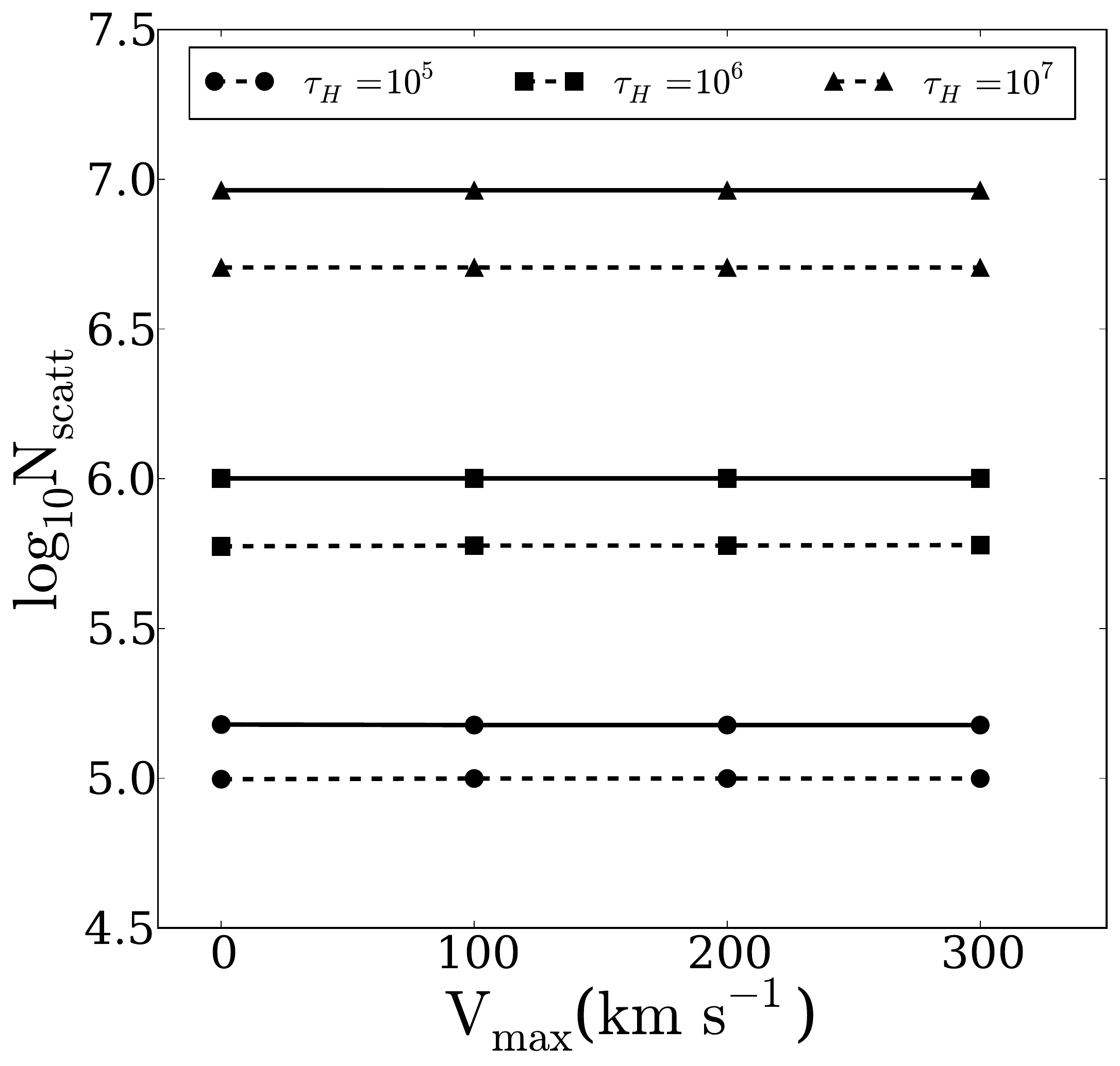}
\end{center}
\caption{2D histogram of the logarithm of the average number of scatterings as function of $\mu$ (left) and the maximum rotational velocity $V_{\rm
max}$ (right). The left panel shows the behaviour for $\tau=10^{5}$ and
  $V_{max}=300$\kms as a function of $\left|\cos\theta\right|$, the color indicates the number of photons per bin. In the
  right panel the continous (dashed) lines represent the results for
  the central (homogeneous) model. The independence of $N_{\rm scatt}$ with $\mu$ and $V_{\rm max}$ is
  present in all models.  
\label{fig:Nscatt} }
\end{figure*}

The number of scatterings affects the escape frequency of a \ly
photon. Studying this quantity further illustrates the independence of
the integrated flux and the escape fraction on rotational velocity.

In Fig. \ref{fig:Nscatt} we show the average number of scatterings
$\langle N_{\rm scatt}\rangle$ as a function of the cosinus of the
outgoing angle $|\cos\theta|$ and the rotational velocity
$V_{\rm max}$. 
From the right panel observe that the number of scatterings and the
outgoing angle are independent. 
This plot corresponds to the specific case of the central model with
$\tau=10^5$ and $V_{\rm max}=300$\kms, but we have verified that this
holds for all models. 

The right panel of Fig. \ref{fig:Nscatt} shows how the average
number of scatterings is also independent from the rotational
velocity. 
The lower number of average scatterings in the homogeneous source
distribution is due to a purely geometrical effect.
Photons emitted close to the surface go through less scatterings
before escaping.

In static configurations it is expected that the optical depth correlates number of
scatterings.  
This has been precisely quantified in the case of static infinite
slab. 
In that model for centrally emitted sources the average number of
scatterings depends only on the optical depth $\langle N_{\rm
  scatt}\rangle=1.612\tau_{\rm   H}$ \citep{Adams72,Harrington73}, for
homogeneously distributed sources $\langle N_{\rm
  scatt}\rangle=1.16\tau_{\rm   H}$ \citep{Harrington73}.

In our case we find that for the central model the number of
scatterings is proportional to the optical depth, with $\langle N_{\rm
  scatt}\rangle= (1.50, 1.00, 0.92)\tau_{\rm   H}$ for optical depth
values of $\tau_{\rm H} = (10^{5}, 10^{6}, 10^{7})$ respectively.
For the homogeneous sources we find that $\langle N_{\rm
  scatt}\rangle= (0.99, 0.59, 0.51)\tau_{\rm   H}$.

\section{Discussion}
\label{sec:discussion}

\subsection{Towards an analytical description}

There is a key result of our simulations that allows us to build an
analytical description for the outgoing spectra.
It is the independence of the following three quantities with the rotational
velocity and the viewing angle: integrated flux, average number of
scatterings and escape fraction.

As we explained in \S~\ref{sec:intlineint}, the best way to understand this is that radiative transfer inside a sphere that undergoes solid-body rotation
proceeds identical to that inside a non-rotating sphere. While scattering events off atoms within the rotating cloud impart
Doppler boosts on the Ly$\alpha$ photon, these Doppler boost are only
there in the lab-frame. Therefore, in the frame of the rotating gas cloud all atoms are
stationary with respect to each other and the scattering process
proceeds identical as in the static case (also see \S~\ref{sec:intlineint} for an additional more quantitative explanation).

This result allows us to analytically estimate the spectrum emerging from a rotating cloud:
The spectrum of \lya photons emerging from a rotating gas cloud is identical as for the static case in a frame that is co-rotating
with the cloud. However, the surface of cloud now moves in the lab-frame. 
Each surface-element on the rotating cloud now has a bulk
velocity with respect to a distant observer. In order to compute the
spectrum one can integrate over all the surface elements in the
sphere with their corresponding shift in velocity and an additional
weight by the surface intensity.

\begin{figure*}
\begin{center}
  \includegraphics[width=0.49\textwidth]{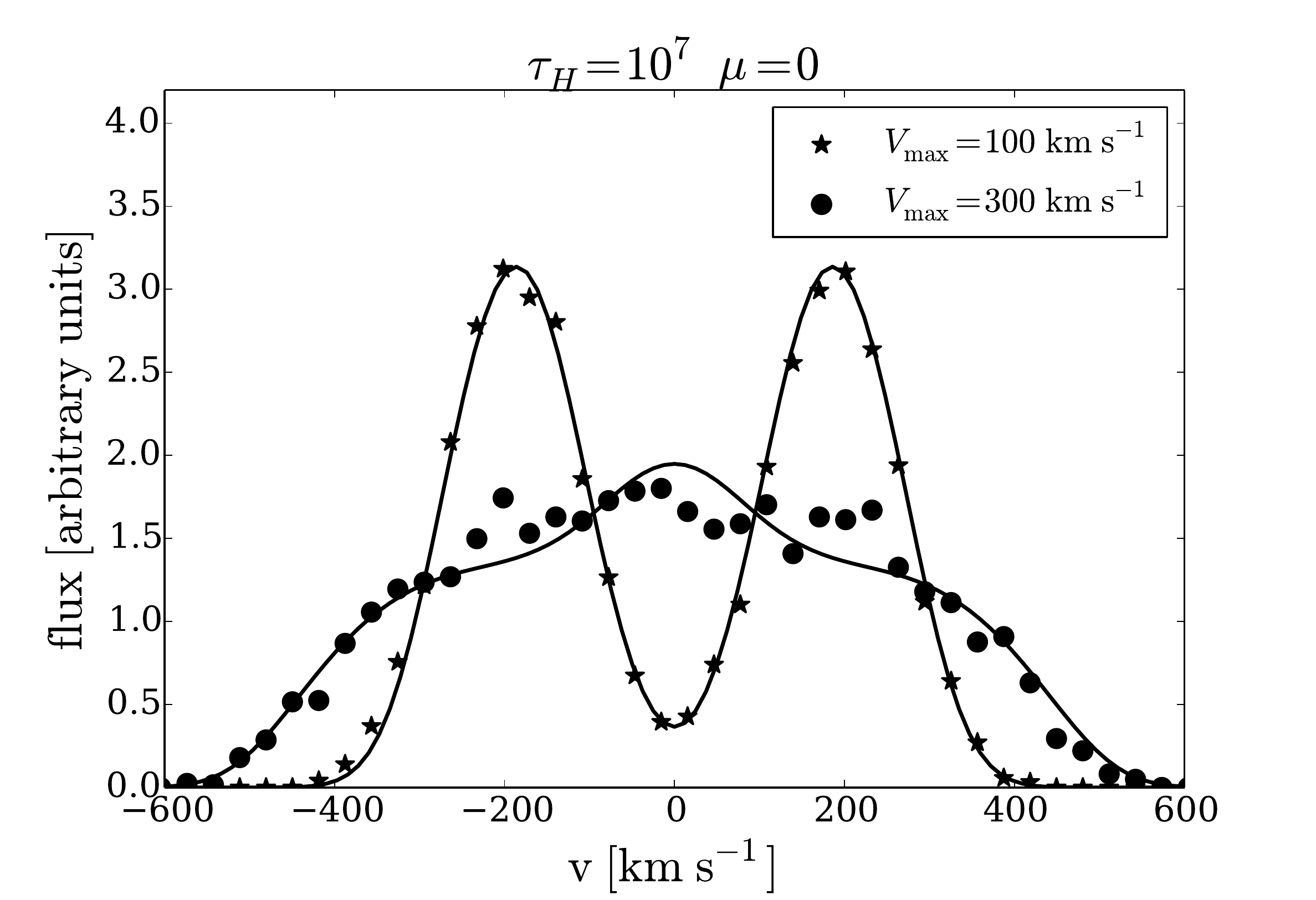}
  \includegraphics[width=0.49\textwidth]{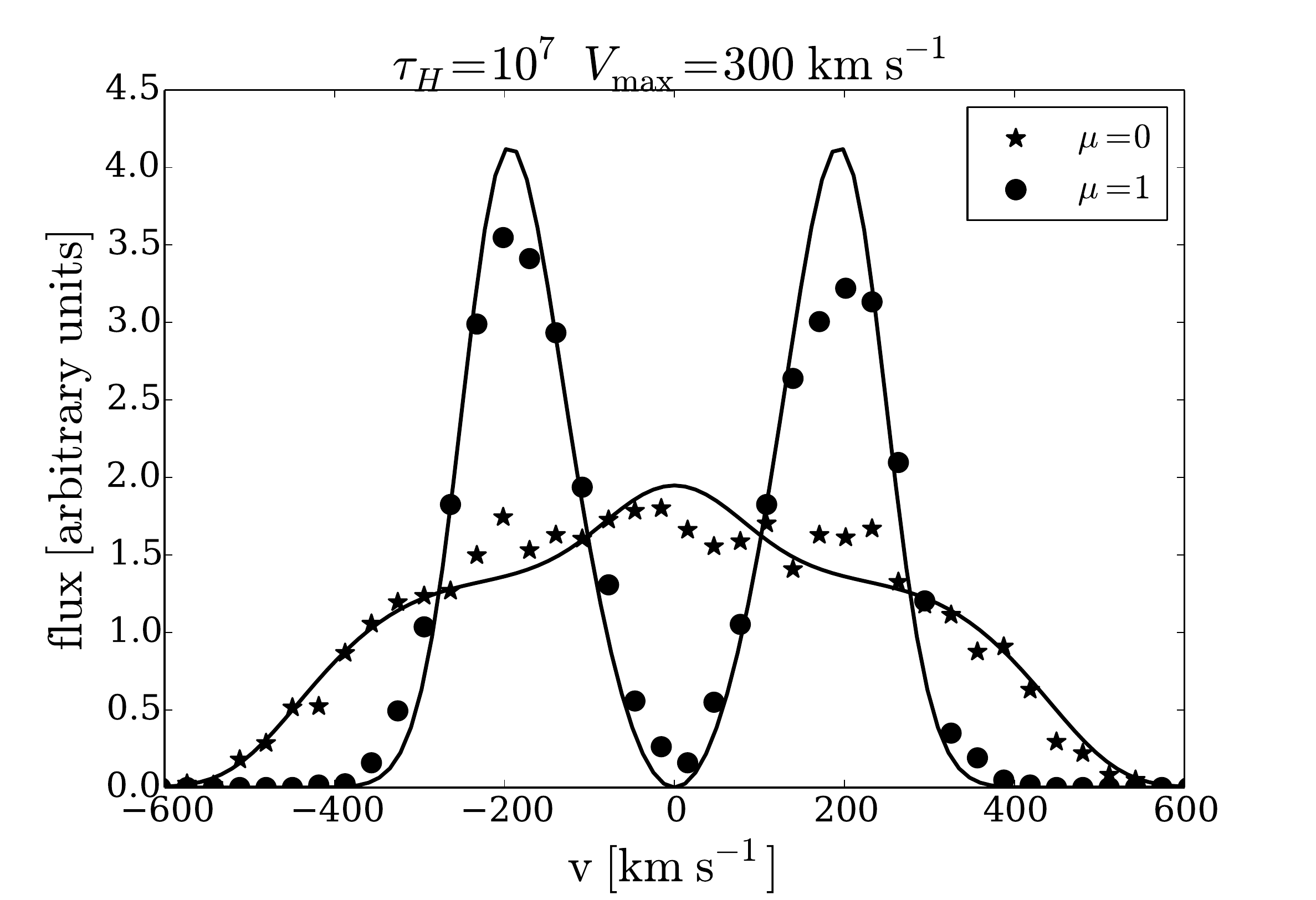}
\end{center}
\caption{
Comparison of the Monte Carlo results against the analytic
solution. The left panel explores the results of different velocities.
The right panel presents the results for two different observers:
paralel and perpendicular to the rotational axis, $\mu=1$ and $\mu=0$
respectively.
\label{fig:comparison} }   
\end{figure*}

Fig~\ref{fig:comparison} shows some examples of analytic versus full MC
spectra using this approach (the implementation details are in the Appendix).
The left panel shows the results for different rotational velocities
in the case of $\tau_{H}=10^7$ and an observer located perpendicular
to the axis of rotation ($i=0$ in the scheme of Fig~\ref{fig:scheme}
in the Appendix). The right panel shows the results for different viewing angles in the
case of $\tau_{H}=10^7$ and a rotational velocity of $V_{\rm
  max}=300$\kms. 

The two methods clearly give good agreement, though not perfect.  In particular, the left panel shows that the MC gives rise to a spectrum that is
slightly more concentrated towards the line centre. As we explain in Appendix~\ref{sec:app}, we do not expect perfect agreement, because this requires an analytic solution for the spectrum of Ly$\alpha$ photons emerging from a static, optically extremely thick cloud {\it as a function of the angle at which they escape from the sphere}. This solution does not exist in the literature. It is possible to get better agreement my modifying the surface brightness profile.
In any case, the analytic calculation closely captures the results
obtained from the full calculations from the MC simulations.  
As such, they are extremely useful and provide us with a quick tool to
verify our calculations at the first order level.

\subsection{Impact on the interpretation of simulated and
  observational data}

We now compare our findings to other computational results and discuss
its possible implications  for the interpretation of observational data.

{\bf Escape at Line Center.} Our models have shown that rotation
enhances the flux density at line center (see Fig.     \ref{fig:differentvelocities}). It has
recently been proposed that galaxies with Lya spectral lines that
contain flux at line center may be `leaking' ionizing (LyC) photons
\citep{Behrens2014,2014arXiv1404.2958V}. The main reason for this possible
connection is that the escape of ionizing (LyC) photons requires
$N_{\rm HI} < 10^{17} $cm$^{-2}$. The same low column densities facilitate the escape of
\ly photons at (or close to) line center. Our work suggests that
rotation may provide an alternative explanation.

{\bf Single peaked lines}. The presence of single peaked profiles has
been associated to inflow/outflow dynamics
\citep{Verhamme06,DijkstraKramer}. 
Gas bulk rotation can also be considered as a probable origin for that
behaviour, provided that the observed single peak is highly
symmetric. 
Similarly, in the case of double peaked lines with a high
level of flux at the line center, rotation also deserves to be
considered in the pool of possible bulk flows responsible for that feature,
specially if the two peaks have similar intensities. 

{\bf Systemic velocities}. There are observational measurements for the
velocity shift between the \ly and other emission lines. In our study
we find that the position of the peak maxima can suddenly change with
rotation and viewing angle. Namely the line can become single peaked
for high rotational velocities and viewing angles perpendicular to the
rotation axis.

{\bf Galaxy simulations with gas rotation}. \cite{Verhamme12} studied \ly
line emission in two high resolution simulations of individual
galaxies. 
The main purpose of their study was to assess the impact of two
different ISM prescriptions. 
However, each simulated galaxy had a disc structure with a clear rotation pattern in
the ISM and inflowing gas from the circum-galactic region. 
The configuration had an axial symmetry and they reported a strong dependence of both
the escape fraction and the total line intensity as a function of the
$\theta$ angle. 
From our study, none of these two quantities has a dependence either
on the inclination angle or the rotational velocity. 
We suggest that he effect reported by \cite{Verhamme12} is
consistent with being a consequence of the different hydrogen optical
depth for different viewing angles and not as an effect of the bulk
rotation.

{\bf Zero impact on the \ly escape fraction}. Study of
high redshift LAEs in numerical simulation often requires the
estimation of the \ly escape fraction in order to compare their
results against observations
\citep{CLARA,Dayal2012,Forero12,Orsi12,Garel2012}. Most of these
models estimate the escape fraction from the column density of dust and
neutral Hydrogen. The results of our simulation indicate that the
rotational velocity does not induce additional uncertainties in those
estimates.

\section{Conclusions}
\label{sec:conclusions}

In this paper we quantified for the first time in the literature the effects
of gas bulk rotation in the morphology of the \ly emission line in
star forming galaxies.   
Our results are based on the study of an homogeneous sphere
of gas with solid body rotation. 
We explore a range of models by varying the rotational speed, hydrogen
optical depth, dust optical depth and initial distribution of \ly
photons with respect to the gas density. 
As a cross-validation, we obtained our results from two independently
developed Monte-Carlo radiative transfer codes.  

Two conclusions stand out from our study. 
First, rotation clearly impacts the \ly line morphology; the width and
the relative intensity of the center of the line and its peaks are
affected. 
Second, rotation introduces an anisotropy for different viewing
angles. 
For viewing angles close to the poles the line is double peaked and it
makes a transition to a single peaked line for high rotational
velocities and viewing angles along the equator. 
This trend is clearer for spheres with homogeneously distributed
radiation sources than it is for central sources.

Remarkably, we find three quantities that are invariant with respect
to the viewing angle and the rotational velocity: the integrated flux,
the escape fraction and the average number of scatterings.
These results helped us to construct the outgoing spectra of a
rotating sphere as a superposition of spectra coming from a static
configuration. This description is useful to describe the main
quantitative features of the Monte Carlo simulations.

Quantitatively, the main results of our study are summarized as
follows. 

\begin{itemize}

\item In all of our models, rotation induces changes in the line morphology
 for different values of the angle between the rotation
 axis and the LoS, $\theta$.  The changes are such that for 
 a viewing angle perpendicular to the
 rotation axis, and high rotational velocities the line becomes single peaked.

\item The line width increases with rotational
  velocity. For a viewing angle perpendicular to the rotation axis
  This change approximately follows the functional form  ${\rm FWHM}^2
  = {\rm FWHM}_{ 0}^2 + (V_{\rm max}/\lambda)^2$, where FWHM$_{0}$
  indicates the line 
  width for the static case and $\lambda$ is a constant. We have
  determined this constant to be  $\lambda_{\rm c}=0.83 \pm 0.06$ and
  $\lambda_{\rm h}=0.82\pm 0.05$ for the central and homogeneous source
  distributions, respectively.

\item At fixed rotational velocity the line width decreases as $|\mu|$
  increases, i.e. the smallest value of the line width is observed for
  a line of sight parallel to the ration axis. 

\item The single peaked line emerges at viewing angles $\mu\sim 1$ for
  when the rotational velocity is close to than half the FWHM$_0$.
\end{itemize}

Comparing our results with recent observed LAEs we find that 
morphological features such as high central line flux, single peak
profiles  could be explained by gas bulk  rotation present in these
LAEs. 

The definitive and clear impact of rotation on the \ly morphology
suggests that this is an effect that should be taken into account at
the moment of interpreting high resolution spectroscopic data. In
particular it is relevant to consider the joint effect of rotation the
and ubiquitous outflows (Remolina-Gutierrez et al., in prep.)
because rotation can lead to enhanced escape of \ly at line center, which
has also been associated with escape of ionizing (LyC) photons
\citep{Behrens2014,2014arXiv1404.2958V}

\section*{Acknowledgments}

JNGC acknowledges financial support from Universidad de los
Andes. 

JEFR acknowledges financial support from Vicerrectoria de
Investigaciones at Universidad de los Andes through a FAPA grant.

We thank the International Summer School on AstroComputing
2012 organized by the University of California High-Performance
AstroComputing Center (UC-HiPACC) for providing computational
resources where some of the calculations were done. 

The data, source code and instructions to
replicate the results of this paper can be found
here {\texttt{https://github.com/jngaravitoc/RotationLyAlpha}}.
Most of our code benefits from the work of the IPython and Matplotlib
communities \citep{IPython,matplotlib}.

We thank the referee for the suggestions that allowed us to greatly
improve and better frame the interpretation of our simulations.

\appendix
\section{Analytic Expression for the Ly$\alpha$ Spectrum
  emerging from Rotating Cloud}  
\label{sec:app}

Ly$\alpha$ scattering through an optically thick gas cloud that is
undergoing solid-body rotation (i.e. in which the angular speed around the
rotation axis is identical for each hydrogen atom) proceeds identical
as in a static cloud. In order to compute the spectrum emerging from a rotating cloud, we sum
the spectra emerging from all surface elements of the cloud, weighted by their intensity.  

\begin{figure*}[h]
\centerline{\includegraphics[width=80mm]{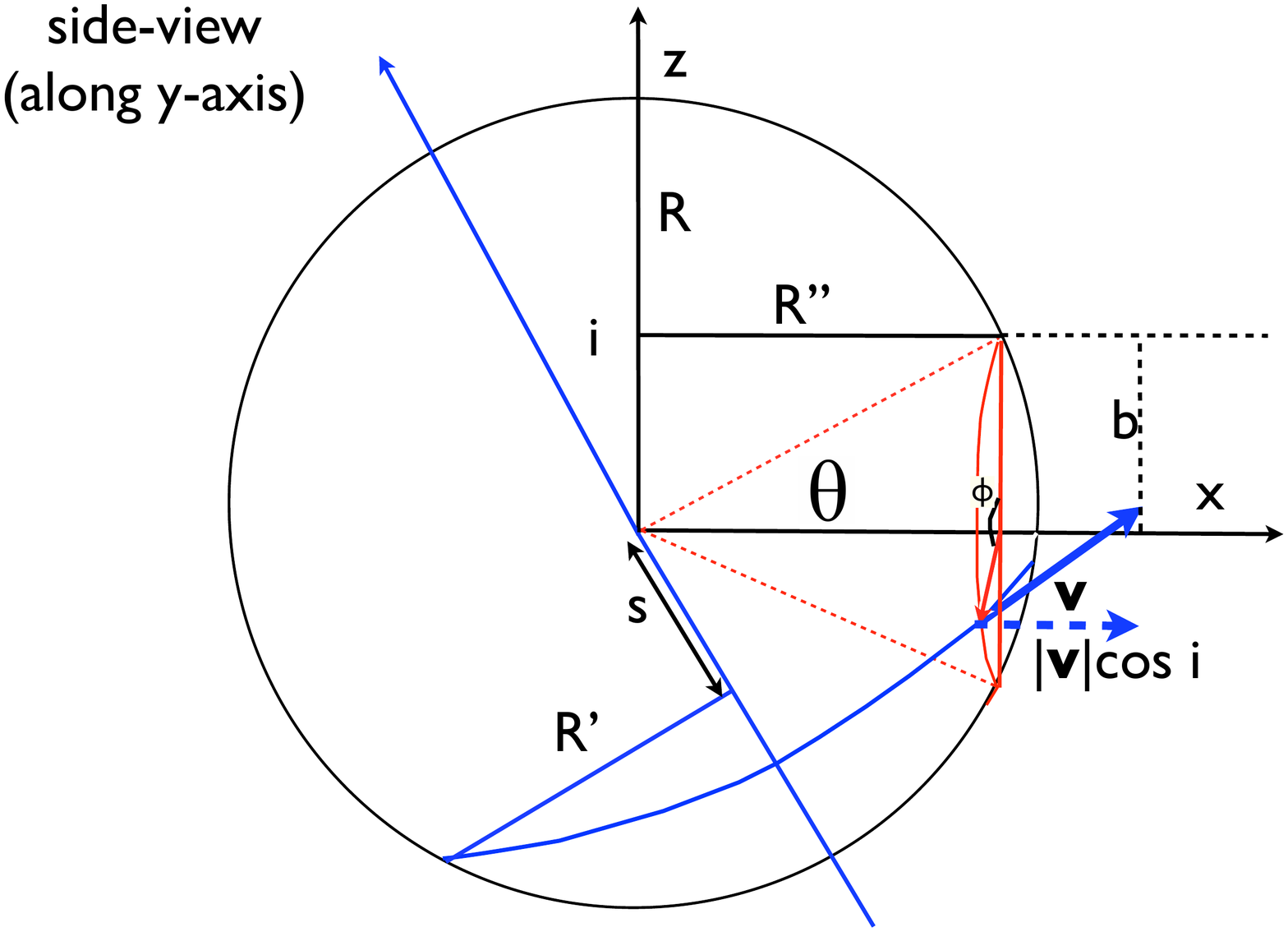}
\includegraphics[width=80mm]{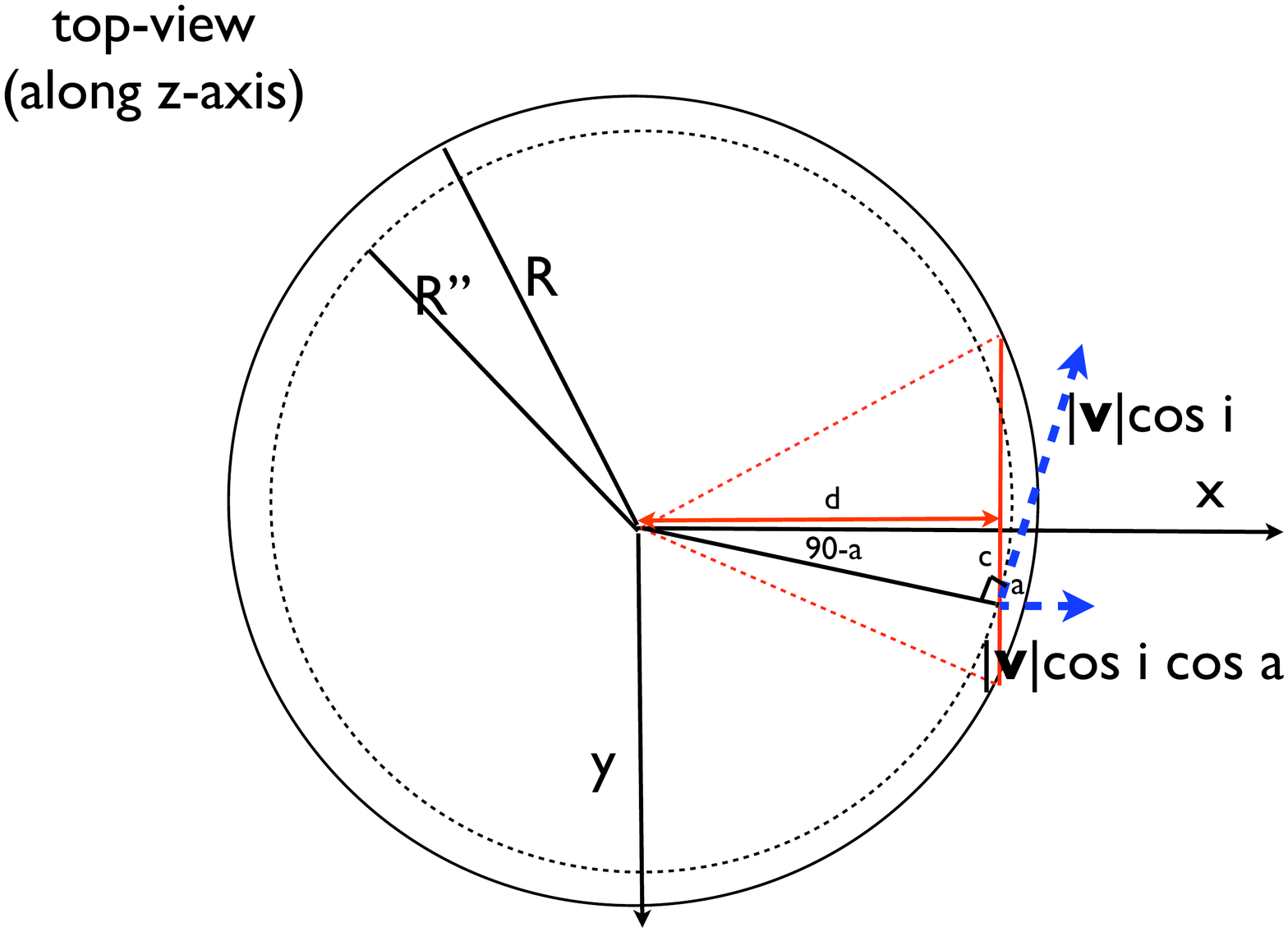}}
\caption[]{Adopted geometry for evaluating the analytic spectrum.}
\label{fig:scheme}
\end{figure*}

We adopt the geometry shown in Fig~\ref{fig:scheme} to derive an analytic expression of this emerging spectrum,
Note that this geometry differs from the scheme shown in Fig~1 in the main body of
the paper. 
The sightline to the observer \& rotation axis define the $x-z$ plane. 
The {\it left panel} in Fig~\ref{fig:scheme} shows the view from the $y$-axis. 
The observer sits along the $x$ axis. 
The rotation axis makes an angle $i$ with respect to the $z$-axis. 
We sum up spectra from individual patches by integrating over the
impact parameter $b$, and angle $\phi$. 
Each $(b,\phi)$ corresponds to a point on the sphere. 
This point has a velocity vector ${\bf v}(b,\phi,i)$, which we denote
with ${\bf v}$ for brevity. The magnitude of ${\bf v}$ is $|{\bf v}|=V_{\rm max}R'/R$. Here $R'=\sqrt{R^2 -s^2}$, in which $s$ denotes the distance of the point $(b,\phi)$ to the plane perpendicular to the rotation axis and through the origin (see the {\it left panel} of Fig~\ref{fig:scheme}). This distance $s$ is given by $s=|-\sin i\sqrt{R^2-b^2}+ b \cos \phi \cos i|$.\\

The spectrum of the flux emerging from the surface at point $(b,\phi)$ is 

\begin{displaymath}
J(x,b,\phi,i)=\frac{\sqrt{\pi}}{\sqrt{24}a\tau_0}\Bigg{(}\frac{(x-x_{\rm
    b})^2}{1+{\rm cosh}\Big{[}\sqrt{\frac{2\pi^3}{27}}\frac{|(x
      -x_{\rm b})^3|}{a\tau_0}\Big{]}}\Bigg{)},
\end{displaymath} 
where $x_b\equiv v_{b}/v_{\rm th}$, and $v_b$ is the component of ${\bf v}$ projected onto the line-of-sight. This component is given by
\begin{equation}
v_{\rm b}(b,\phi,i)=V_{\rm max}\frac{\sqrt{R^2 -s^2}}{R}\cos i  \hs
\cos a,
\end{equation}
where $\beta = 90^{\circ}-a$. The factor $\cos i$ accounts for the projection onto the $x-y$ plane, and the factor $\cos a$ for the subsequent projection onto the line-of-sight. The {\it right panel} of Fig.~\ref{fig:scheme} shows that this angle $a$ can be computed from
\begin{equation}
\tan \beta =\tan[90^{\circ}-a]=\frac{c}{d}=\frac{ b\sin \phi}{\sqrt{R^2 -b^2}},
\end{equation}

In order to compute to total intensity we integrate over $b$ and
$\phi$ with a weight given by the surface brightness of the
sphere at $(b,\phi)$, $S(b,\phi)$.

\begin{displaymath}
J(x,i)=2\pi \int_0^Rdb \hs b \int_0^{2\pi}d\phi \hs
S(b,\phi)J(x,b,\phi,i) \approx 2\pi \int_0^Rdb \hs b
\int_0^{2\pi}d\phi \hs J(x,b,\phi,i)\\ \nonumber.
\end{displaymath}
In the last expression we assume that $S(b,\phi)$ is constant. 
This corresponds to $I(\mu) \propto \mu$ at the surface, where $\mu$
denotes the cosine of the angle of the propagation direction of the
outgoing photon and the normal to the spheres surface: a fixed $db$
corresponds to a physical length $ds = db/\mu$ on the sphere.  
If $I(\mu)$ were constant, this would imply that the sphere should
appear brighter per unit $b$. 
A constant surface brightness profile requires the directional
dependence for $I(\mu) \propto \mu$ to correct for this. 
Indeed, this is what is expected for the escape of Ly$\alpha$ photons
from static, extremely opaque media (see \citet{Ahn01}; their
Fig~4 and accompanying discussion). 

It is worth stressing that this derivation should not be viewed as a
complete analytic calculation, and we do not expect perfect agreement:
we {\it assumed} a functional form for the surface brightness profile
[or for $I(\mu)$]. Moreover, $I(\mu)$ itself may depend on frequency $x$. In other words, analytic solutions exist for $J(x) =
\int_0^1 I(x,\mu) d\mu$ at the boundary of the sphere, and {\it approximate}
expressions for $I(\mu) =\int dx I(x,\mu)$, but {\it not} for $I(x,\mu)$
itself. The spectra we obtained from the Monte-Carlo calculations naturally include the proper $I(x,\mu)$, and are therefore expected to be more accurate.

\begin{figure*}[h]
\centerline{
\includegraphics[width=0.30\textwidth]{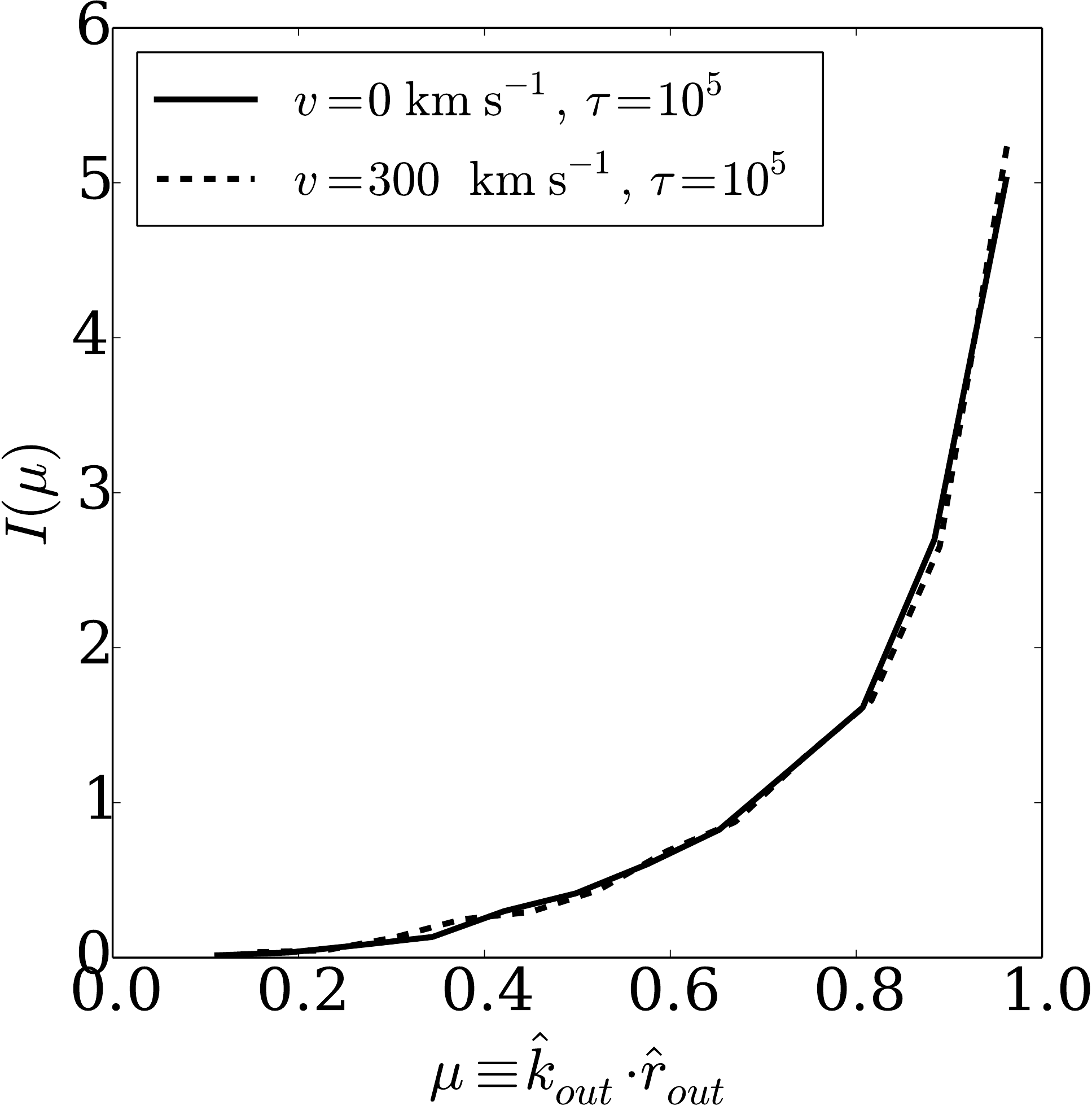}
\includegraphics[width=0.30\textwidth]{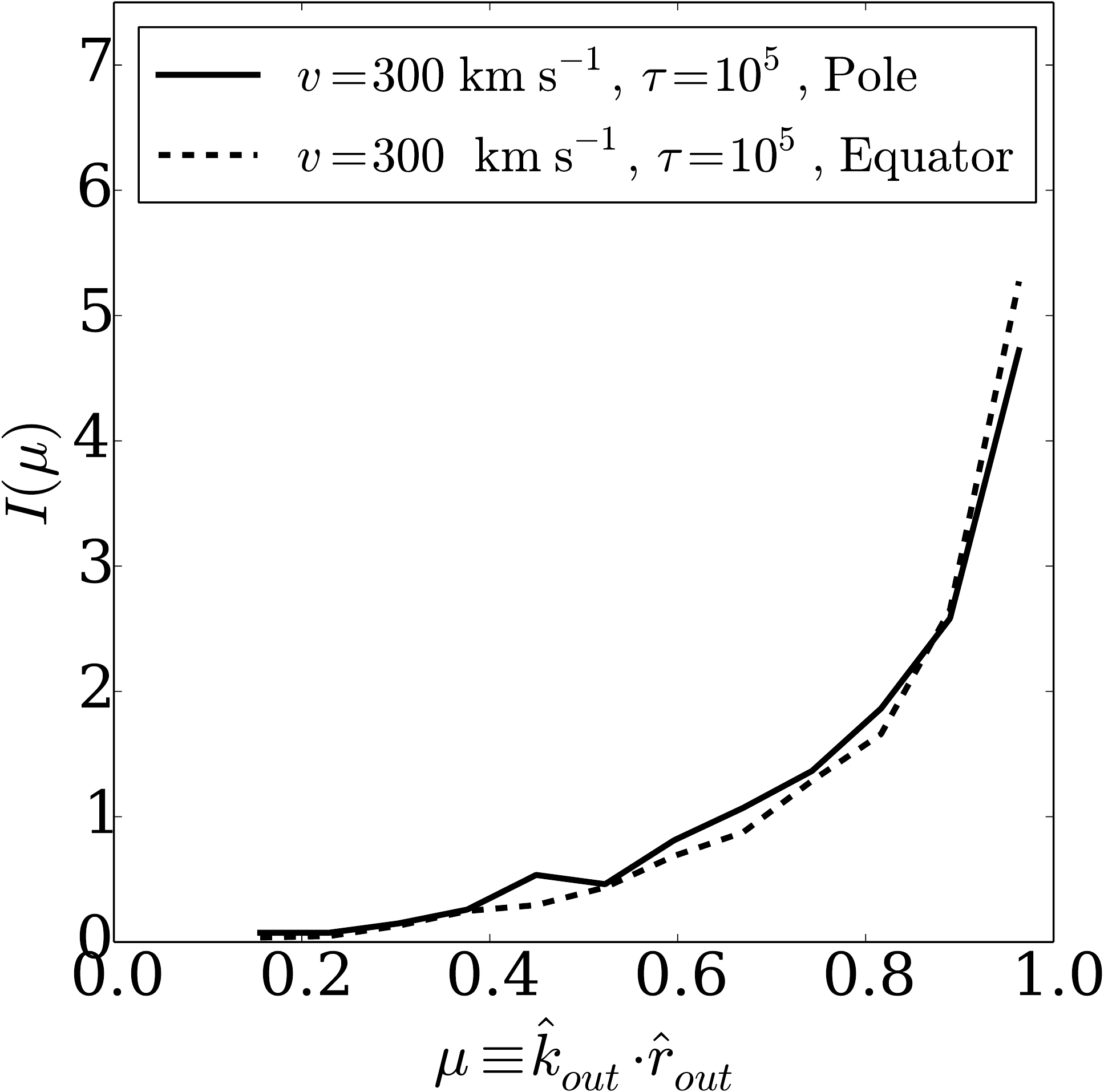}
\includegraphics[width=0.30\textwidth]{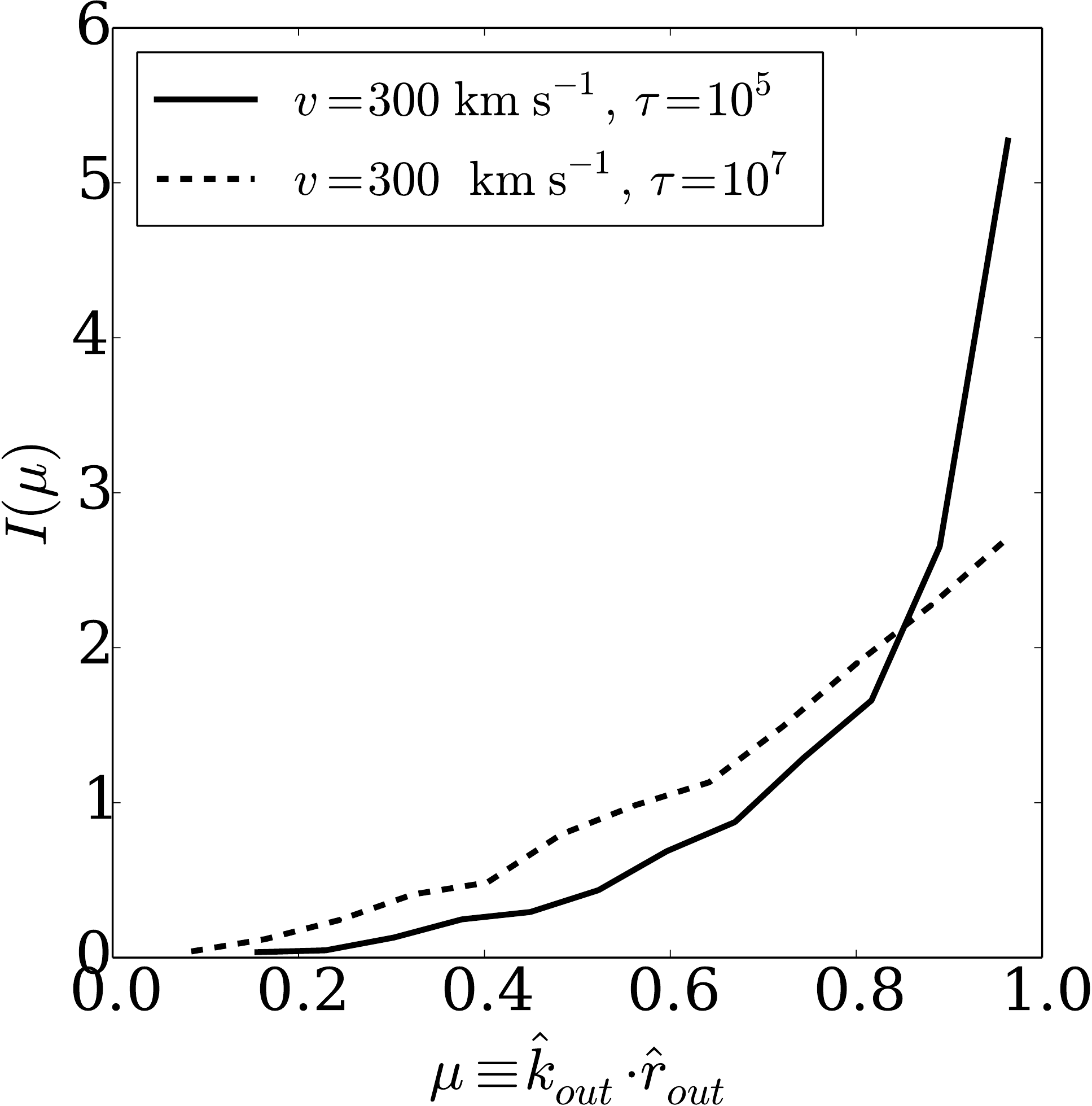}}
\caption[]{Distribution of the cosine of the angle between the
  propagation direction and a vector normal to the sphere's
  surface. The distributions have been normalized to unity. Left
  panel: different rotational velocities; middle panel: different
  viewing angles; right panel: different optical depths. Only the
  optical depth has an effect on the distribution of outogoing
  directions. This is consistent with the assumption that \lya
  scattering in a medium with solid body rotation proceeds as in a
  static medium.} 
\label{fig:surface}
\end{figure*}

To further test the assumption of scattering in a rotating medium proceeding
as in a static medium we compute the distribution of the outgoing
angles $\mu$. The results are shown in Figure \ref{fig:surface}; it
shows that the distribution for $\mu$ is independent of the rotational
velocity and the location over the sphere. The only dependence comes
with $\tau_{H}$. For higher values of the optical depth the
distribution gets closer to $I(\mu)\propto \mu$ as expected for a
static medium \citep{Ahn01}.

\bibliographystyle{apj}
%\bibliography{references} 

\end{document}